# Neural networks in pulsed dipolar spectroscopy: a practical guide

Jake Keeley[1], Tajwar Choudhury[1], Laura Galazzo[2], Enrica Bordignon[2],
Akiva Feintuch[3], Daniella Goldfarb[3], Hannah Russell[4], Michael J. Taylor[4],
Janet E. Lovett[4], Andrea Eggeling[5], Luis Fabregas Ibanez[5], Katharina Keller[5],
Maxim Yulikov[5], Gunnar Jeschke[5], Ilya Kuprov[1,*]

[1]School of Chemistry, University of Southampton, Southampton SO17 1BJ, United Kingdom.
[2]Department of Physical Chemistry, University of Geneva,
Quai Ernest Ansermet 30, CH-1211 Geneva, Switzerland.
[3]Department of Chemical Physics, Weizmann Institute of Science, Rehovot 7610001, Israel.
[4]SUPA School of Physics and Astronomy, and BSRC, University of St Andrews,
North Haugh, St Andrews, KY16 9SS, United Kingdom.
[5]Department of Chemistry and Applied Biosciences, Swiss Federal Institute of
Technology in Zurich, Vladimir Prelog Weg 2, CH-8093 Zürich, Switzerland.

*i.kuprov@soton.ac.uk

## Abstract

This is a methodological guide to the use of deep neural networks in the processing of pulsed dipolar spectroscopy (PDS) data encountered in structural biology, organic photovoltaics, photosynthesis research, and other domains featuring long-lived radical pairs and paramagnetic metal ions. PDS uses distance dependence of magnetic dipolar interactions; measuring a single well-defined distance is straightforward, but extracting distance distributions is a hard and mathematically ill-posed problem requiring careful regularisation and background fitting. Neural networks do this exceptionally well, but their "robust black box" reputation hides the complexity of their design and training – particularly when the training dataset is effectively infinite. The objective of this paper is to give insight into training against simulated databases, to discuss network architecture choices, to describe options for handling DEER (double electron-electron resonance) and RIDME (relaxation-induced dipolar modulation enhancement) experiments, and to provide a practical data processing flowchart.

# 1. Introduction

The recent evolution of statistical analysis and numerical regression into machine learning and neural nets has been a remarkable success. The reason has been known since 1980s: some neural networks are universal approximators [1,2] – but it was not before teraflop scale computing power had arrived that highly visible applications emerged. One of those is deconvolution, defined formally as solving Fredholm equations [3]. Neural nets are apparently able to learn the kernel and produce numerically stable inversions of the convolution operator [4,5]. Applications include image and voice recognition [6-8], deeply subwavelength optical microscopy [9], magnetic resonance imaging [10], and – in our present case – pulsed dipolar spectroscopy (PDS) of electron spin pairs.

The energy of the magnetic dipole interaction between unpaired electrons depends on the inverse cube of the distance. Electron magnetic moments can be determined precisely, and the resulting distance measurement techniques are called pulsed dipolar spectroscopy [11,12]. Double electron-electron resonance (DEER) is the most popular one; it is essentially a spin echo experiment, modified to keep only the dipole term in the effective Hamiltonian, and implemented so as to avoid the dead times of microwave electronics [13,14]. A similar experiment, called relaxation-induced dipolar modulation enhancement (RIDME) uses electron spin relaxation as a replacement for one of the magnetisation inversion steps [15,16]. Both methods are popular in structural biology because they return distributions of distances and therefore offer a window into nanometre-scale conformational mobility [17,18]; RIDME is particularly useful with paramagnetic metal ions [19,20]. When unpaired electrons are not present naturally (*e.g.* in metalloproteins), they are introduced by site-directed spin labelling: most commonly, selected amino acids are mutated into cysteines and stable radicals are attached to the thiol sulphur [21]. The most popular spin label is *S*-(1-oxyl-2,2,5,5-tetramethyl-2,5-dihydro-1H-pyrrol-3-yl)methyl methanesulphonothioate (MTSL). It is highly reactive towards cysteines; it is also small and flexible enough to avoid disturbing protein folds [22].

In rigid and precisely oriented molecules, distances between 15 and 80 Å are easily measured with high accuracy [23], but in soft matter the problem is more difficult. Firstly, a distribution of distances would normally be present. Secondly, biological samples rarely have directional order – experimental data is an average over all orientations. Thirdly, the presence of out-of-pair interactions creates a background signal that depends on placement topology and concentration of external spins [24]. All of this, and the inevitable presence of noise, makes PDS data processing an ill-posed problem in the Hadamard sense [25]: in the absence of regularisation, the solution is neither unique nor a smooth function of the experimental data [26]. Mathematically, the difficulty is somewhere between the inverse Fourier transform [27] and the inverse Laplace transform [28] – not as easy as the former, but not as woefully unstable as the latter. In common with other deconvolution problems, neural networks do well [29] for reasons that are only starting to be explored [30].

A recent application of pulsed dipolar spectroscopy is nanometre-scale distance measurement performed in frozen biological cells [31-33], in particular using Gd(III) spin labels [34-37] that are chemically stable in the cytoplasm. Gd(III) tags produce an easily detectable ESR signal because their $(-1/2) \to (+1/2)$ electron spin transition has a narrow line at W-band and above [38]. They are also useful in mixed labelling methods where the partner spin comes from an organic radical, such as nitroxide [39,40] or trityl [41]. However, the complicated quantum dynamics in Gd(III)-Gd(III) spin systems (S=7/2, large zero-field splitting) yields uninterpretable data unless certain safety margins are observed that allow the use of the weak coupling approximation [42,43] – pulse frequencies must be chosen so that the system stays in the weak dipole-dipole coupling regime [44,45]. Within those margins, Gd(III) spin labels are useful and informative [34,36-41], but the general form of Gd(III)-Gd(III) DEER kernel is unknown. A similar situation exists in RIDME spectroscopy where high-frequency overtones are present in high-spin systems, meaning that no simple analytical form exists for the kernel function [46]. This brings us back to neural networks that, from a large enough collection of examples, can learn the kernel (Figure 1) in a way that does not depend on magnetic or structural parameters of any specific physical system – the only requirements are: (a) knowing the mathematical class of distance distributions that can be encountered in PDS measurements; (b) being able to construct a training set that prepares the neural network for dealing with any distribution from that class.

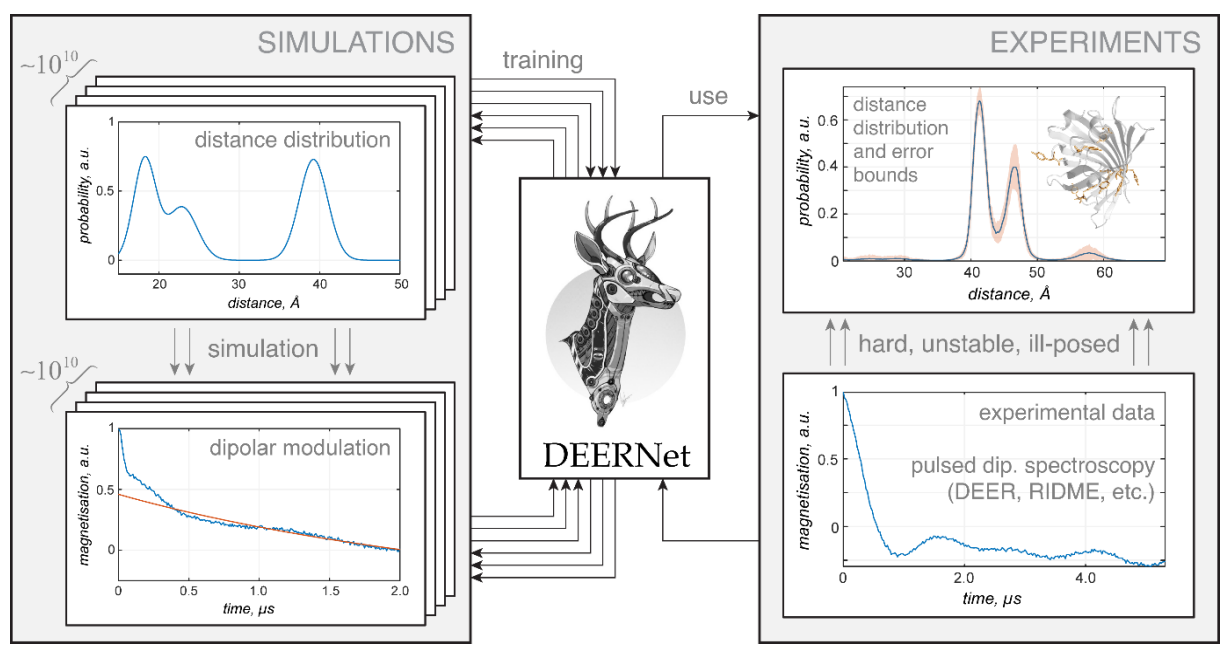

*Figure 1. DEERNet training and inference flowchart. The block schematic shows the relationship between the computationally generated training database, the neural network, and its practical application in the context of pulsed dipolar spectroscopy (DEER, RIDME, etc.) distance measurement. To converge the training process, a carefully designed database with many billions of question-answer pairs is necessary; it can only be generated by highly accurate simulations including models of instrumental artefacts and noise. At the point when the network is used, it has no adjustable parameters.*

An attractive feature of a trained neural net is the absence of user-adjustable parameters. Existing DEER and RIDME data processing tools feature sophisticated regularisation [47,48] and model fitting [49] algorithms with many unobvious settings, including the choice of background model and uncertainty estimation, each with its own parameters. Inexperienced users picking unsuitable parameter combinations has long been acknowledged as a problem. DEERNet does not expose any adjustable parameters to the user, yet its performance matches existing regularisation-based tools [29,50]. Unusually for machine learning, the important question about *exactly how* the neural network produces the output signal has been answered [30] – DEERNet appears to be a combination of digital filters and regularised integral transforms. It is not fully transparent, but it is translucent.

The utility of neural networks for solving Fredholm equations [3] that arise in pulsed dipolar spectroscopy is firmly established [5,29,30]; the challenges at this point are logistical: network architecture, training database generation, convergence of the training process, uncertainty estimation, processing sparsely sampled data, and enforcing applicability for the approximations and assumptions made in the underlying models. These logistical aspects are the focus of the present paper.

## 2. DEERNet architecture

DEERNet is a signal processing network trained to solve the following inhomogeneous Fredholm equations of the second kind for the distance distribution function $p(r)$:

$$\left[(1-\mu)+\mu\int p(r)\gamma(r,t)dr\right]b(t)+n(t)=s(t)$$
$$b_{\text{DEER}}(t)=\exp\left[(-kt)^{N/3}\right] \qquad b_{\text{RIDME}}(t)=\exp\left[-a_1 t-a_2 t^2\right] \tag{1}$$

where $\gamma(r,t)$ is a known kernel [48,51], $s(t)$ is the experimentally recorded signal [51], $n(t)$ is Gaussian white noise with unknown standard deviation, $k$ is an unknown DEER background decay rate [11], $N$ is an unknown DEER background dimension [24], $a_{1,2}$ are unknown RIDME background decay parameters [52], and $\mu$ is an unknown modulation depth [11,51]. The problem is exceedingly difficult [26], but much progress was made over the last twenty years [47,48,51]. Neural networks were found to be surprisingly effective [29] – a recent in-depth inspection revealed digital filters that isolate the integral in Eq (1) followed by spectral filtering regularisation as a means of inverting the convolution operation [30].

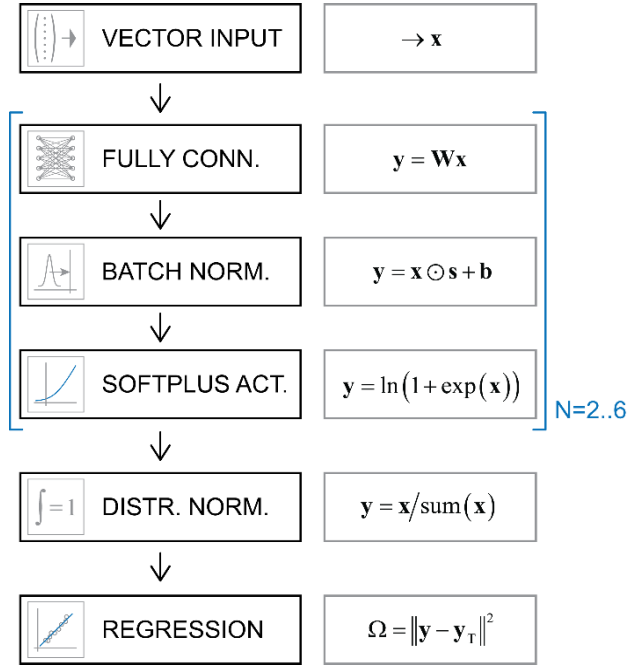

*Figure 2. Architecture of the second generation DEERNet. The blocks on the right show the mathematical operations performed by the corresponding layers. The adjustable parameters are weight matrices **W** in the fully connected layers, as well as scale and shift vectors **s** and **b** in the batch normalisation layers. The strong performance relative to the state of the art [29] is due to the fact that the mathematical operations in question are close to those needed in a Fredholm solver [4,5].*

The architecture of the original DEERNet [29,30] did not respect the physical requirement for asymptotic linearity of the DEER data processing problem – all activation functions were sigmoidal or logsigmoidal. The first generation network also made no use of convergence acceleration tools such as batch normalisation [53], and had no internal provision for normalising the output, which has a physical meaning of probability density. These matters are rectified in the second-generation architecture shown in Figure 2 and implemented in *Spinach 2.6* and later [54]. The new DEERNet is a feed-forward neural network with repeating triads of fully connected (FC), batch normalisation (BN), and softplus activation (SA) layers, terminated by a probability distribution normalisation layer.

Each FC-BN-SA triad applies the following transformation to the input vector $\mathbf{x}$:

$$\mathbf{y} = \ln\left\{\exp\left[(\mathbf{W}\mathbf{x})\odot\mathbf{s}+\mathbf{b}\right]+1\right\} \qquad (2)$$

where $\odot$ stands for element-wise multiplication. This equation and the flowchart in Figure 2 warrant an extended explanation. The matrix-vector multiplication $\mathbf{W}\mathbf{x}$ is a linear transformation that – in the absence of instabilities and distortions – the DEER processing problem should have been. The subsequent scaling by $\mathbf{s}$ and shift by $\mathbf{b}$ are also a linear transformation – it is called batch normalisation [53]; it accounts for statistical variations between training data batches, thereby improving the convergence of the training process (Figure 3, top left). The remaining $\mathbf{y} = \ln\left\{\exp[\ldots]+1\right\}$ transformation, called softplus [55], is the non-linearity required by the universal approximator theorem [1,2] – without it, the network could be rearranged into a single linear operation. Softplus function is non-linear at the origin, but asymptotically linear:

$$\ln\{\exp[x]+1\} = x + O(x^{-1}), \quad x \gg 1 \qquad (3)$$

At the last FC-BN-SA triad in Figure 2, the choice of softplus activation function was dictated by the nature of DEERNet output – probability density, a non-negative quantity without an upper bound. At the intermediate triads, softplus appears empirically (not shown) to be a better choice than the more popular logsigmoidal function; this is likely because the network as a whole approximates an unstable but linear transformation. The number of FC-BN-SA blocks in Figure 2 is set to six because there is no improvement beyond that (Figure 3, bottom row). The final distribution normalisation (DN) layer enforces the physical requirement for all probabilities to sum up to 1; its presence yields a significant performance boost (Figure 3, top right), particularly for narrow distance distributions. It also makes the output independent of the modulation depth parameter $\mu$ in Equation (1).

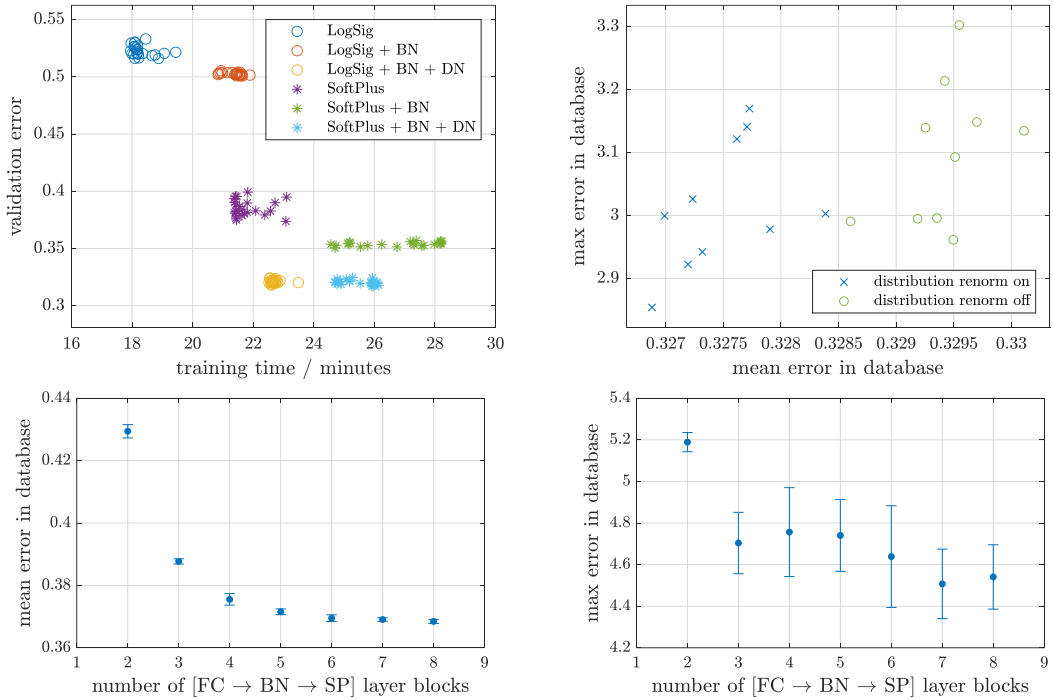

*Figure 3. Performance effect of neural network architecture decisions. Scatter plots and statistics refer to sets of independently trained (different random initial guess, different databases) DEERNets on a batch of 64,000 datasets generated as described in [29] and Section 3 below. The error is defined as root mean square deviation between the network output and the ground truth for the 512-element non-negative output vector, normalised (for numerical accuracy reasons in single-precision arithmetic) to have the mean value of 1. **(Top left)** Validation error and training time for different choices of activation function (logsigmoidal vs. softplus [55]) in the presence or absence of batch normalisation (BN) layers [53], and distribution normalisation (DN) layers. **(Top right)** Effect of the distribution normalisation layer on the mean and maximum (over the test database) regression error by the networks. **(Bottom row)** Mean and maximum error (over the test database) as functions of the number of [fully connected]-[batch normalisation]-[softplus activation] triads in the architecture shown in Figure 2.*

## 3. Infinite training database

The number of parameters in a typical DEERNet (Figure 2) is in the millions – there is not enough real experimental DEER or RIDME data in existence to train a network of this size. However, accurate sim-ulations [48,54] are computationally affordable, and decades of experimental work have yielded practical intervals for the parameters of those simulations. Second-generation DEERNets in *Spinach 2.6* and later are trained on simulated signal databases with the parameters listed in Table 1.

DEER background parameters are well understood [11], but the choice of RIDME background parameter ranges in Table 1 merits a discussion, because it is empirical: $b_{\text{RIDME}}(t)$ in Eq (1) only applies rigorously for $T_{1E}/T_{2E} > 10$ and no distribution of $T_{1E}$ times in the sample – elsewhere, it merely has the status of something that fits the data well (here, $T_{1E}$ refers to the pumped spin and $T_{2E}$ refers to the detected spin [11]). Neural network training databases must be representative of the experimental data that the network would later see, but the definition of "representative" is subjective. From a recent theoretical treatment [52] and from practical experience [20], it is reasonable to assume that, if a RIDME background function is initially steady or growing, that it would have reached a turning point by $t = t_{\max}/5$ where $t_{\max}$ is the duration of the RIDME trace. We therefore seek such a combination of parameters $a_{1,2}$ as would guarantee that:

$$\begin{cases} \left[\dfrac{\partial}{\partial t}\exp\left(-a_1 t - a_2 t^2\right)\right]_{t=0} \geq 0 \\ \left[\dfrac{\partial}{\partial t}\exp\left(-a_1 t - a_2 t^2\right)\right]_{t \leq t_{\max}/5} = 0 \end{cases} \Rightarrow \begin{cases} a_1 \leq 0 \\ a_2 \geq -5a_1/2t_{\max} \end{cases} \quad (4)$$

Practical experience also indicates that, if a RIDME background function is initially decreasing, that it must continue decreasing for the duration of the experiment:

$$\begin{cases} \left[\dfrac{\partial}{\partial t}\exp\left(-a_1 t - a_2 t^2\right)\right]_{t=0} < 0 \\ \left[\dfrac{\partial}{\partial t}\exp\left(-a_1 t - a_2 t^2\right)\right]_{t=t_{\max}} < 0 \end{cases} \Rightarrow \begin{cases} a_1 > 0 \\ a_2 > -a_1/2t_{\max} \end{cases} \quad (5)$$

Physically, $a_1$ is the rate constant associated with the loss of spin correlation; users may be instructed to truncate the data set to a time point $t_{\max}$ when the dipolar modulation is no longer visible, corresponding roughly to $a_1 = 3 t_{\max}^{-1}$ – we therefore chose this to be the standard deviation of the normal distribution of the $a_1$ parameter in the training database (Table 1). A similar extent of endpoint decay with respect to the $a_2$ parameter is achieved for $a_2 = 3 t_{\max}^{-2}$, but its distribution is also bounded by the inequalities in Eqs (4) and (5). Accordingly, a half-normal distribution is used for $a_2$, from those bounds out into the positive direction (Table 1).

This kind of loose reasoning invites criticism; its justification comes from the nature of the background elimination process in DEERNet, which is known to be a digital filter [30] that simply destroys the background. Accordingly, we need only show the network the general class of signal components which it must learn to destroy without quantification – we need not be rigorous.

*Table 1. Parameters used in the generation of training databases. Where ranges are given, a value is picked randomly with the specified probabilities (discrete quantities) or from the specified distributions (continuous quantities) for each input-output data pair in the database. Variables refer to Equation (1).*

| Parameter | Values |
|---|---|
| Network input and output vector dimension | 512 |
| Number of distance peaks in the distribution | 1 - 3 (equal probability) |
| Full width at half-magnitude for distance peaks | from 5% of the distance to 50% of the distance range |
| Background decay rate $k$, fraction of $t_{max}^{-1}$ | 0.0 - 1.0 (uniform distribution) |
| Background dimension $N$ | 2.0 - 3.5 (uniform distribution) |
| RIDME background parameter $a_1$ | normal distribution $\langle a_1 \rangle = 0, \quad \sigma[a_1] = 3t_{max}^{-1}$ |
| RIDME background parameter $a_2$ | half-normal distribution $\langle a_2 \rangle = \max\{-2a_1/5, -a_1/2\}$ $\sigma[a_2] = 3t_{max}^{-2}$ |
| Modulation depth $\mu$ | 0.01 - 1.00 (uniform distribution) |
| Standard deviation of the noise in $n(t)$ | $[0.00 - 0.05] \cdot \mu$ (uniform distribution) |

Although the maximum number of distance peaks in the training database is set to a sensible value of three, the resulting networks can apparently handle more [29]. Asymmetric distance peak shapes are also reproduced (Supplementary Information, Section S2) even though the training database only contains Gaussians [29,30]. This stability to excursions outside the training range is unusual in machine learning; it may be the consequence of the unique appropriateness of deep networks with fully connected layers specifically for the deconvolution problem – rigorous convergence bounds exist [5] for approximating solutions of integral equations with neural nets, and Fredholm equations were among the earliest applications [4]. Apparently, networks are learning to actually solve Eq (1) as opposed to memorising or interpolating solutions [30].

Since the data source is artificial (Figure 1), the size of the training database is theoretically infinite. This removes the complications associated with overfitting – it is possible to design a database (in this case, a *Matlab* datastore object) that serves previously unseen data every time a batch is requested by the stochastic gradient descent algorithm [56], meaning that each converged network has effectively been trained on an infinite database. This approach has a useful side effect of reducing random-access memory and disk storage requirements: by the time the training process has converged, a typical DEERNet will have seen a few billion question-answer pairs – terabytes of data. Storing that in a pre-computed form would have been impractical. A minor logistical optimisation is that the calculation of data batches may be done outside the thread that runs the training. Annotated source code and API documentation for all of the above is supplied with *Spinach* (http://spindynamics.org).

## 4. Background and modulation depth

Distance distribution is not the only information contained in the time-domain DEER data: the background function and the modulation depth are also useful [57-59]. However, because both are destroyed within the neural network [30], their extraction must follow a roundabout path: first the probability density is converted back into the form factor, then the form factor is mixed with an analytical background model, and then the mixture is retrofitted to the time-domain data. The stages are:

1. The normalised distance distribution is converted into the form factor $f(t)$ using the standard DEER kernel [26,47]. This is a stable and therefore unproblematic transformation:

$$f(t) = \int p(r)\gamma(r,t)dr \quad \Leftrightarrow \quad \mathbf{f} = \mathbf{\Gamma p} \tag{6}$$

where $\mathbf{f}$ and $\mathbf{p}$ are a discrete representations of $f(t)$ and $p(r)$ on finite grids, and $\mathbf{\Gamma}$ is the corresponding matrix representation of $\gamma(r,t)$.

2. The parameters of the background model are extracted from a least squares fit of the noiseless DEER signal model in Eq (1) to the experimental data:

$$\begin{aligned}\{\alpha,\mu,k,N\} &= \arg\min \left\| s_{\text{expt}}(t) - s_{\text{theo}}(t) \right\|^2 \\ s_{\text{theo}}(t) &= \alpha\left[(1-\mu) + \mu f(t)\right]b(t)\end{aligned} \tag{7}$$

where $\alpha$ is a multiplier reflecting the overall scaling of the input signal, $\mu$ is the modulation depth, and $b(t)$ is a background model. First generation DEERNet [29] had a separate network set for background extraction, but the retrofitting approach advocated here appears to be superior in our practical testing. It also enables processing of batches of signals that share a common background; this feature is implemented in the new version.

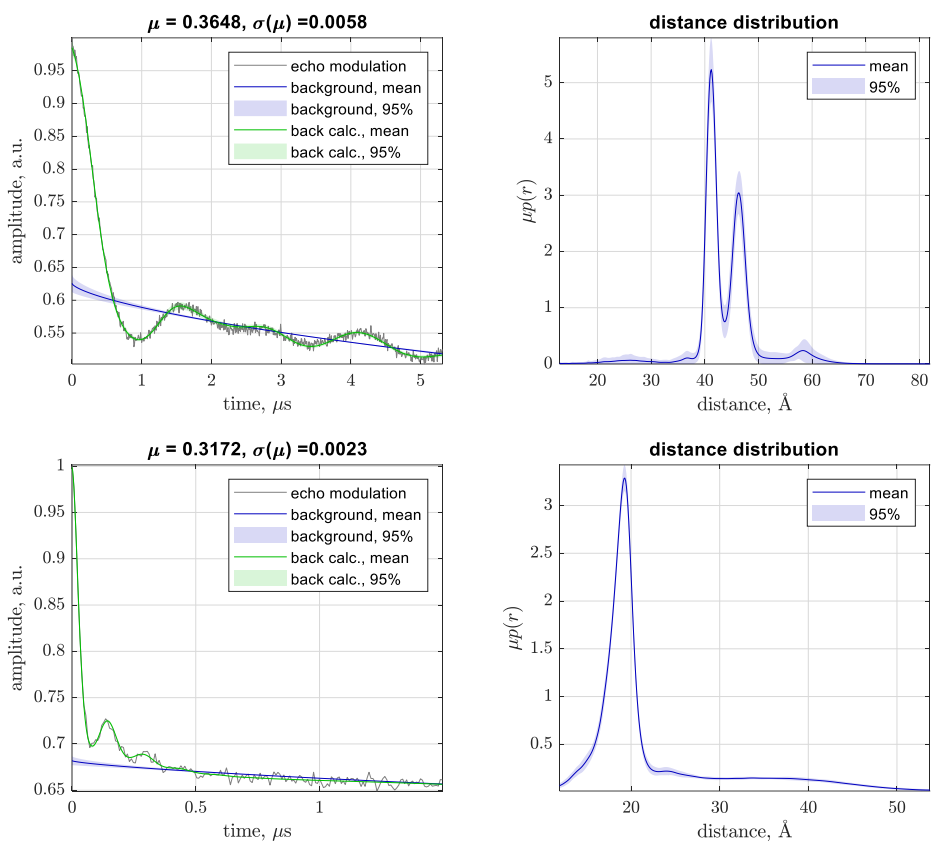

*Figure 4. Distance distribution and background extraction performance. The illustrations match those of the previous generation DEERNet reported in Figures 11 and 15 of [29]; note the improved distance range. **(Top row)** V96C/I143C site pair in the lumenal loop of a double mutant of light harvesting complex II, using iodocateamido-PROXYL spin labels attached to the indicated cysteines [60]. Residue 96 is located in the lumenal loop, and residue 143 is a structurally rigid "anchor" position in the protein core. In agreement with the results reported in the original paper [60], a bimodal distance distribution is obtained, indicating flexibility in the lumenal loop. The low-confidence peak around 57 Angstrom likely results from protein aggregation. **(Bottom row)** Pairs of nitroxide radicals tethered to the surface of gold nanoparticles, with the thiol tether attachment points diffusing on the surface of the nanoparticle (sample Au 3 after solvolysis and heating in [61]). The broad pedestal is real, it matches the analytical model reported in [61]; existing regularisation methods cannot process this dataset without either broadening the peak or introducing clumping artefacts into the pedestal.*

## 5. Uncertainty quantification

Output uncertainty analysis in DEER spectroscopy has been extensively researched [48-51,62,63], but neural nets offer two more methods beyond what is normally considered – the network ensemble method and the linear uncertainty propagation. The former is unique to neural networks and the latter proceeds from the fact that, unlike the formal convolution inversion operation, a neural network is stable (in the sense of having finite derivatives) with respect to its inputs.

### 5.1 Network ensemble method

In the machine learning community a popular way of estimating the uncertainty in the neural network output is to train an ensemble of nets on different databases from a different random initial guess, and to run descriptive statistics on their outputs [29] – examples (shaded blue intervals) are given in the right hand panels of Figure 4 and Supplementary Information figures. Training 32 networks appears empirically to be sufficient; it is important to note that we did not try to ruggedise the networks against deliberately crafted adversarial inputs [64].

There are two sources of uncertainty in the distance distribution data: one associated with the distance distribution extraction process itself, and the other associated with the modulation depth obtained in Stage 2. These uncertainties are statistically independent because the probability density is normalised, meaning that the first point of the form factor is always exactly 1. The uncertainties of the outputs are therefore obtained by running statistics over sets of products $\{\mu^{(k)} p^{(k)}(r)\}$, $\{s_{\text{theo}}^{(k)}(t)\}$, and $\{(1-\mu^{(k)})b^{(k)}(t)\}$ obtained from the network ensemble. The former yields the shaded confidence intervals in the right panels of Figure 4, and the latter two produce the blue (background) and the green (overall signal) shaded confidence intervals in the left panels.

## 5.2 Linear uncertainty propagation

A vector-in, vector-out neutral net $\mathbf{y} = \Upsilon(\mathbf{x})$ with continuous activation functions is easy to differentiate with respect to the input vector; this may be done by a finite difference approximation [65] or using the same automatic differentiation methods that compute the gradient for the training process [66]. The derivative of the output vector with respect to the input vector is the Jacobian matrix $\mathbf{J}$:

$$\mathbf{y} = \Upsilon(\mathbf{x}), \quad J_{nk} = \partial y_n / \partial x_k \tag{8}$$

The appearance of the network Jacobian (Figure 5, left panel), relative to the Jacobian obtained by differentiating Eq (1) analytically (Figure 5, right panel), illustrates of the stability of the neural net with respect to perturbations of its input. The network is apparently only sensitive to variations in the input data when those variations influence significant distance peaks (see the top row of Figure 4 for the distance distribution). The values of the derivatives are moderate throughout.

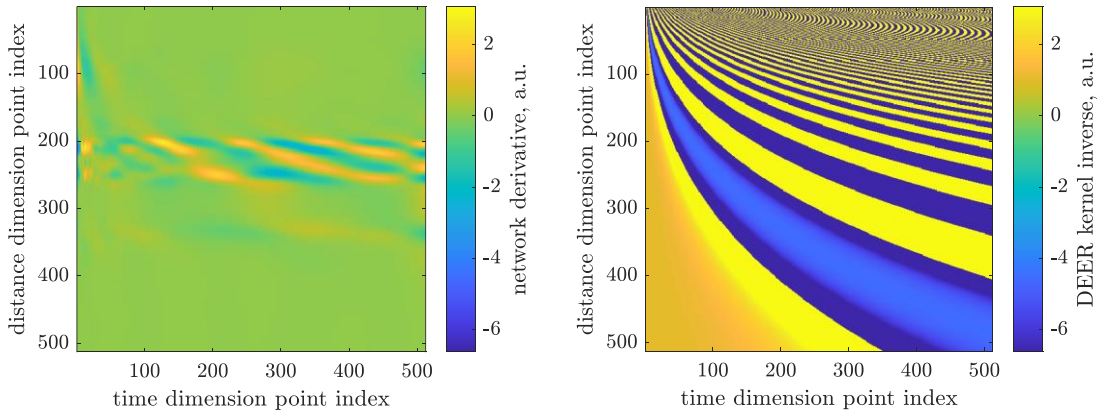

*Figure 5. (Left)* A colour map representation of the Jacobian matrix of DEERNet (an average over 32 networks trained on different databases from a different random initial guess) for the input-output pair appearing in the top row of Figure 4. The horizontal dimension corresponds to the input vector and the vertical dimension to the output vector. *(Right)* A clipped (to the same range as the left panel) colour map representation Jacobian matrix obtained by computing the point-by-point inverse of the DEER kernel in Eq (9).

That is emphatically not the case for the Jacobian obtained from Eq (1), where the statistical expectation value of the variational derivative of the distance distribution with respect to the input signal has entire curves filled with infinities because the kernel $\gamma(r,t)$ is zero on those curves:

$$\langle \delta p(r)/\delta s(t) \rangle = \left[ \mu \gamma(r,t) b(t) \right]^{-1} \qquad (9)$$

This is of course the exact reason why conventional processing of DEER data requires regularisation; it would seem that the neural network has found a way around the problem. At the moment, there is only limited understanding of how it does that [30]; on the bright side, the availability of first derivatives enables linear uncertainty propagation:

$$\sigma_{y_n}^2 \approx \sum_k \left( \frac{\partial y_n}{\partial x_k} \right)^2 \sigma_{x_k}^2 = \sum_k J_{nk}^2 \sigma_{x_k}^2 \qquad (10)$$

in which standard deviations $\sigma_{x_k}$ may be estimated from the fitting residuals in the time domain, or from the difference between the raw signal and a suitably filtered (for example, by the Savitsky-Golay procedure) version. This is illustrated for a RIDME dataset in Figure 6.

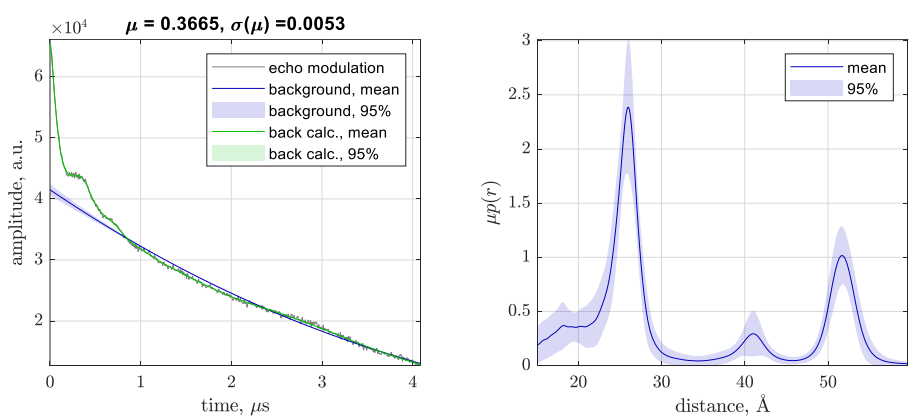

*Figure 6. An example of linear uncertainty propagation in the distance distribution.* The sample is a protein homodimer of copper amine oxidase from Arthrobacter globiformis (AGAO) containing a $Cu^{2+}$ ion bound to a surface site on each monomer and one MTSL spin label per monomer. RIDME data was collected with a mixing time of 5 µs. Two dominant $Cu^{2+}$ to MTSL distances are expected: one intra-monomer (26 Å) and one intra-dimer (52 Å). Two-stage regularised processing methods that require background elimination report zero confidence for the biologically certain 52 Å distance (Section S4 in the SI) – single-stage processing with neural networks is clearly superior in this case. Sample preparation procedures and details of data collection may be found in Figure 3a of [67].

## 6. Sparsely sampled data

Because DEER and RIDME signals are detected as indirectly incremented dimensions of two-dimensional ESR experiments, there is much scope for instrument time savings using the same sparse sampling methods that are popular in NMR spectroscopy [68-70]. However, an attempt to train a neural net by feeding sparsely sampled data as index-value pairs predictably fails. We also abstain from existing solutions that work well [71], but without – hopefully not for long – an explanation of how.

The key to obtaining interpretable and effectual networks in this case is what we call the "no magic" assumption – based on current evidence [30], we expect digital signal processing networks to invent efficient implementations of known mathematics. Given a sparsely sampled input, we could expect a fitting or an interpolation procedure to emerge inside; the input should therefore be adjusted to make it easier for the network to do that.

Polynomial interpolation and fitting share the same mathematical structure – a low-order polynomial is assumed to describe the data well in a sufficiently small interval:

$$f(t) = a_0 + a_1 t + a_2 t^2 + ... \quad (11)$$

and its coefficients are obtained by solving, exactly or approximately, the system of equations requiring the polynomial to return values $\{f_n\}$ at locations $\{t_n\}$:

$$\begin{pmatrix} 1 & t_1 & t_1^2 & \cdots \\ 1 & t_2 & t_2^2 & \cdots \\ 1 & t_3 & t_3^2 & \cdots \\ \vdots & \vdots & \vdots & \ddots \end{pmatrix} \begin{pmatrix} a_0 \\ a_1 \\ a_2 \\ \vdots \end{pmatrix} = \begin{pmatrix} f_1 \\ f_2 \\ f_3 \\ \vdots \end{pmatrix} \quad (12)$$

This system may be overdetermined, in which case it is solved for the coefficients $\{a_k\}$ in the least squares sense. Those coefficients are linear functions of the known values $\{f_n\}$ and rational functions of the time grid locations $\{t_n\}$.

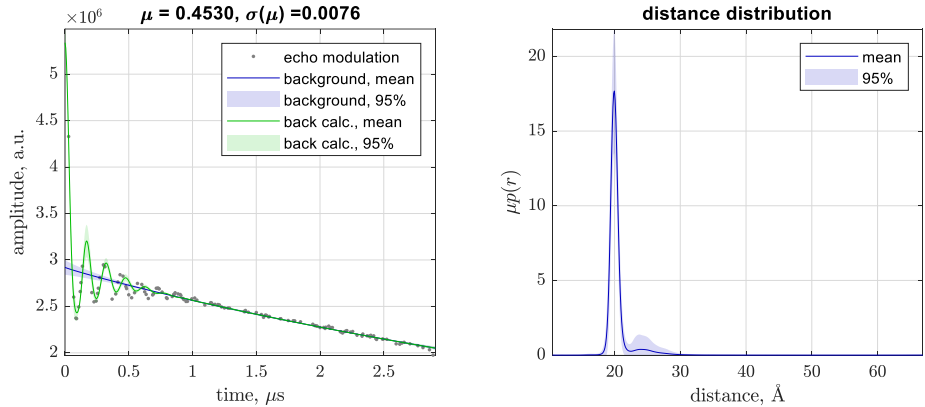

*Figure 7. A sparsely sampled DEER dataset and the result of its neural network processing.* Experimental data come from an X-band DEER measurement of model biradical 3 from Ref [26] diluted into o-terphenyl. The dataset contains 512 time grid points, with the probability of a sample being recorded at each point set uniformly to 1/4.

The internal mechanics of feed-forward neural networks makes it easy for them to learn unary and additive binary functions of the input (sum, inverse, square, *etc.*), but hard to learn multiplicative binary functions – therefore, sets of products like $\{t_n f_n\}$ and $\{t_n^2 f_n\}$ must be supplied as inputs alongside $\{t_n\}$ and $\{f_n\}$. When this is done, network performance approaches the fully sampled case (Figure 7). We found it unnecessary to pick any specific class of sampling schedules [70] at the training stage, but some performance improvements would likely result if sampling schedules are matched between the experimental data and the training database, at the cost of loss of generality.

The only constraint on sparse sampling DEERNets is related to the fixed element count in the input and output vectors of the feed-forward architecture shown in Figure 2: each network ensemble is trained for a specific number of non-empty samples. 128/512 and 64/512 networks are included with DEERNet in *Spinach 2.6*, as are scripts to train network sets for other time grid sizes and sample counts. Although individual data samples may be missing, the underlying time grid must be uniform to enable the calculation of the distance window in the output.

Sparse sampling raises the delicate question of the minimum and maximum distance that can be reliably expected from the distribution reconstruction process. DEER traces are not sinusoidal, but the frequency spectrum of the oscillatory part of the kernel [29]:

$$\gamma(r,t) = \sqrt{\frac{\pi}{6\omega_D t}} \left[ \cos(\omega_D t) \text{FrC}\left(\sqrt{\frac{6\omega_D t}{\pi}}\right) + \sin(\omega_D t) \text{FrS}\left(\sqrt{\frac{6\omega_D t}{\pi}}\right) \right]$$

$$\omega_D = \frac{\mu_0}{4\pi} \frac{\gamma_1 \gamma_2 \hbar}{r^3}; \quad \text{FrC}(x) = \int_0^x \cos(\pi t^2/2) dt; \quad \text{FrS}(x) = \int_0^x \sin(\pi t^2/2) dt \quad (13)$$

is inside $[-2\omega_D, 2\omega_D]$, and therefore – by Whittaker's interpolation formula [72] – the minimum distance captured by a uniformly sampled DEER trace with time grid spacing $\Delta t$ is:

$$\Delta t = \frac{\pi}{2\omega_D^{\max}} \quad \Rightarrow \quad r_{\min}^{\text{DEER}} = \sqrt[3]{\frac{\gamma_1 \gamma_2 \mu_0 \hbar}{2\pi^2} \Delta t} \quad (14)$$

In the uniformly sampled DEERNet, the grid spacing is either that of the original input data, or that of the resampled signal that is supplied to the network, whichever is larger. When individual samples may be missing, no perfect reconstruction conditions are available, but an optimistic choice for $\Delta t$ is the smallest interval seen between the samples that are present.

The maximum distance is set by the requirement to sample at least half a period of $\omega_D^{\min}$. For a uniformly sampled DEER trace with the last sample at $t_{\max}$:

$$t_{\max} = \frac{\pi}{\omega_D^{\min}} \quad \Rightarrow \quad r_{\max}^{\text{DEER}} = \sqrt[3]{\frac{\gamma_1 \gamma_2 \mu_0 \hbar}{4\pi^2} t_{\max}} \quad (15)$$

Here too, when individual samples may be missing, no perfect reconstruction conditions are available, but an optimistic choice for $t_{\max}$ is the time of the last sample that is recorded. All of this arithmetic is performed internally by DEERNet; it is not visible to the user. Note the sliding multiplicative relationship between $\gamma_{1,2}$, $r^3$, and $t$ in the $\omega_D t$ product in Eq (13) – variations in experiment duration and magnetogyric ratios are accommodated by distance axis rescaling; this is handled outside the neural nets which internally use scaled and normalised dimensionless units.

The safety margins must be tighter for RIDME data because RIDME background in Eq (1) is not necessarily monotonic and the assumption of single electron $T_1$ time does not necessarily hold – thus, a full period of the sinusoidal wave is now required for reliably telling dipolar oscillations from the background. Likewise, the short-distance limit of RIDME may contain overtone frequencies in high-spin systems, with the resulting need to sample (and therefore prevent reflections of) frequencies about a factor of two higher than $2\omega_D$. The resulting safety margins are:

$$r_{\min}^{\text{RIDME}} = \sqrt[3]{\frac{\gamma_1 \gamma_2 \mu_0 \hbar}{\pi^2} \Delta t}, \qquad r_{\max}^{\text{RIDME}} = \sqrt[3]{\frac{\gamma_1 \gamma_2 \mu_0 \hbar}{8\pi^2} t_{\max}} \qquad (16)$$

The intervals in Eqs (14)-(16) are used for training database generation and for deciding output distance extents in second-generation DEERNet.

## 7. Practical flowchart

This section provides practical advice on the use of second-generation DEERNet. The package is distributed with *Spinach 2.6* [54] and later (http://spindynamics.org). It is open-source, but requires *Matlab R2021a* or later, with *Optimisation Toolbox*, *Parallel Computing Toolbox*, *Deep Learning Toolbox*, and *Reinforcement Learning Toolbox* installed.

### 7.1 Data preprocessing

The responsibility for supplying valid data rests with the user – internal validation only catches glaring errors (array dimension and type mismatch, unphysical values, *etc.*). The input must be:

(a) Sampled, sparsely or fully, on a uniform time grid. This is dictated by the distance edge mathematics in Eqs (14)-(16), which is only unambiguous for uniform time grids. Data array dimension must be the same as that of the time grid array. Data values corresponding to the missing samples must be set to NaN (not-a-number).

(b) Phased as near to zero imaginary part as the signal-to-noise ratio permits. This may be done manually, by picking such $\varphi$ as to minimise a norm of $\text{Im}\left[e^{i\varphi} s(t)\right]$, or automatically – for example by numerical minimisation of its norm-square:

$$\varphi = \arg\min \left\| \text{Im}\left[e^{i\varphi} s(t)\right] \right\|^2 \qquad (17)$$

Data that cannot be phased in this way is corrupted; it must not be used.

(c) Cropped and shifted so as to make the echo modulation maximum correspond to the $t = 0$ point of the time grid. DEERNet assumes time grid units to be seconds.

Further processing may be required for DEER and RIDME data recoded with non-standard pulse sequences. DEERNet assumes that the data came from (or has been transformed to correspond to) the standard 4-pulse DEER [73] or a 5-pulse RIDME [16] experiment. An input filter for the popular BES3T format is provided.

### 7.2 Input and output data structures

The first argument of the DEERNet function call is a real-valued column vector (or a matrix made of multiple column vectors) containing PDS data; the second argument is a real-valued column vector containing the time grid, and the third argument specifies the experiment ('deer' or 'ridme'):

```
dataset=deernet(input_traces,input_axis,expt);
```

Because feed-forward neural networks have a fixed input dimension and normalisation expectations, appropriate transformations are applied internally; those are automatic and not visible to the user.

The output is a graph with the appearance and content of Figures 4, 6, 7, and a data structure with the fields specified in Table 2. As a matter of policy, DEERNet has no adjustable parameters.

*Table 2. Second-generation DEERNet output. The program returns a Matlab data structure with the indicated fields.*

| Field of the output structure | Content |
|---|---|
| `.input_axis` | time grid (seconds), as received |
| `.input_traces` | DEER or RIDME data, as received |
| `.n_networks` | number of neural nets in the ensemble |
| `.train_params` | neural network training parameters, described in [29] |
| `.resamp_axis`<br>`.resamp_traces` | time grid and data, resampled to 512 time points to match the input dimension of the neural networks |
| `.dist_ax` | distance grid (Angstrom) of the output |
| `.dist_av`<br>`.dist_lb`<br>`.dist_ub` | mean, 95% lower bound, and 95% upper bound on the distance distribution(s); this corresponds to $p(r)$ in Eq (1) |
| `.backgs_av`<br>`.backgs_lb`<br>`.backgs_ub` | mean, 95% lower bound, and 95% upper bound on the background signals; this corresponds to $(1-\mu)b(r)$ in Eq (1) |
| `.retros_av`<br>`.retros_lb`<br>`.retros_ub` | mean, 95% lower bound, and 95% upper bound on the back-calculated fit to the experimental data; this corresponds to $s(t)-n(t)$ in Eq (1) |
| `.mdpths_av`<br>`.mdpths_st` | mean and standard deviation of the modulation depths; this corresponds to $\mu$ in Eq (1) |

## 8. Practical applicability considerations

This section deals with the practicalities of running a regularised version of an unstable mathematical operation; these matters are the same for neural networks and conventional DEER/RIDME data processing methods. The absence of user-adjustable parameters in *DEERNet* is a significant advantage, but the principle of *"garbage in, garbage out"* does of course still apply.

### 8.1 Distances at the long edge

The maximum distance bound in Eqs (15) and (16) ignores the presence of the damping multiplier on the left of the square bracket in Eq (13), of the modulation depth transformation in the square bracket of Eq (7), and of the decaying functions in the baseline term. For this reason, all DEER and RIDME processing methods struggle to tell apart slow dipolar modulations and baseline decays. As a result, distance distributions with significant density at the long edge are unreliable (Figure 8). Because baseline decay is usually significant, the ghost peaks at the long edge of the distance distribution may be reported with high confidence (Figure 8, top right). Those long-distance peaks may sometimes be real (there are examples in the Supplementary Information), but may just as well be artefacts – not a comfortable situation in the data processing context.

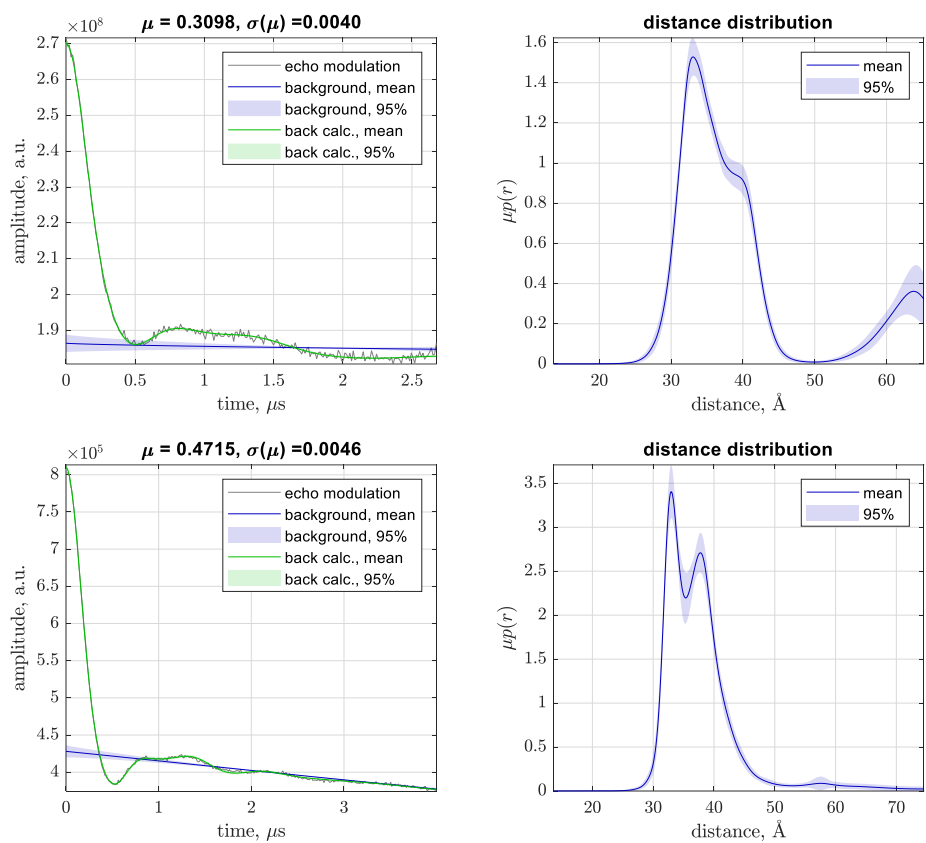

*Figure 8. Example of an artefact peak at the long distance edge. The artefact peak at 65 Å in the top right panel disappears and the true distances become better resolved when a longer DEER trace (bottom row) for the same sample is recorded and processed. Note the incorrect background and therefore incorrect modulation depth in the top panels. Experimental data from Figure S11b of the recent community white paper on PDS data processing [74].*

In such cases, second generation *DEERNet* prints a warning message encouraging the user to record a longer trace, and turns off the output of background signal and modulation depth. This is also done when unreasonable standard deviations are detected in the uncertainty quantification described in Section 5. An appealing feature of *DEERNet* is that background signal retrofitting is optional. This is because backgrounds – with whatever parameters – are eliminated by the notch filter in the first layer of the distance distribution extraction networks [30].

Because RIDME backgrounds are less well understood than DEER backgrounds, two-stage (background elimination followed by deconvolution) Tikhonov-regularised processing methods struggle with longer distances in RIDME data (see the note in the caption of Figure 6). Thus, even in DEERAnalysis2021, the recommended data processing option for such RIDME data is "DEERNet/RIDME". This is further illustrated in Section S4 of the Supplementary Information.

## 8.2 Low modulation depth

A conceptually difficult scenario for any DEER/RIDME processing method is quantifying distances that do not exist – a sample need not contain any radical pairs. Modulation depth quantification described in Section 5 acts as a safeguard here; it is implemented as two criteria: if the modulation depth returned by any of the networks in the ensemble is negative, or if the ensemble averaged modulation depth is below 0.5%, second generation *DEERNet* refuses to output the distance distribution and prints a deliberately forceful error message (Figure 9, right panel). The message is essential: open source software technically allows users to turn safeguards off; it is important to discourage that practice.

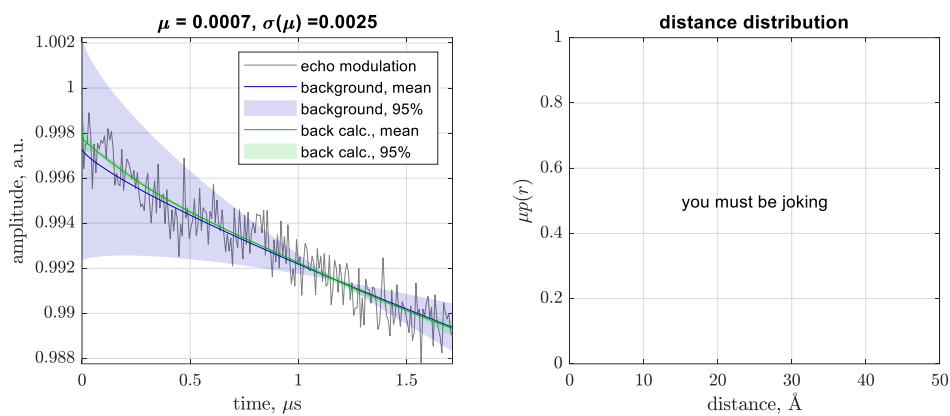

*Figure 9. An illustration of the built-in sanity control in the second-generation DEERNet. Distance distribution output is turned off when data is judged by the neural networks to not constrain it sufficiently well. The left panel shows a Q-band DEER measurement of a singly Gd(III) DOTA maleimide labelled protein ligand in the absence of the binding partner and therefore a priori absence of spin echo envelope modulation. The artefact seen in the left panel is due to microwave pulse interference that is not fully suppressed by phase cycling.*

### 8.3 Running echoes and artefacts

The low modulation depth of Gd(III)-Gd(III) data makes it susceptible to microwave pulse interference artefacts that manifest as running echoes that cross the integration window and may survive phase cycling if observer and pump channel are phase-locked [75]. These running echoes are a common sight in RIDME data; they can appear even in Q-band DEER using Gaussian pulses on nitroxides. Another type of distortion arises at ends DEER time traces due to the residual "2+1" pulse train electron spin echoes [76]; those may sometimes be suppressed using Gaussian pulses with non-overlapping excitation profiles, but may also be unavoidable [77].

A surprising observation is that neural networks are – for want of a better term – ignoring these artefacts (Figure 10). This appears to be a beneficial side effect of the digital filter that they evolve in the input layer [30] – running echoes do not pass through that filter. Should the need arise, running echoes may be introduced as procedurally generated artefacts into the training database.

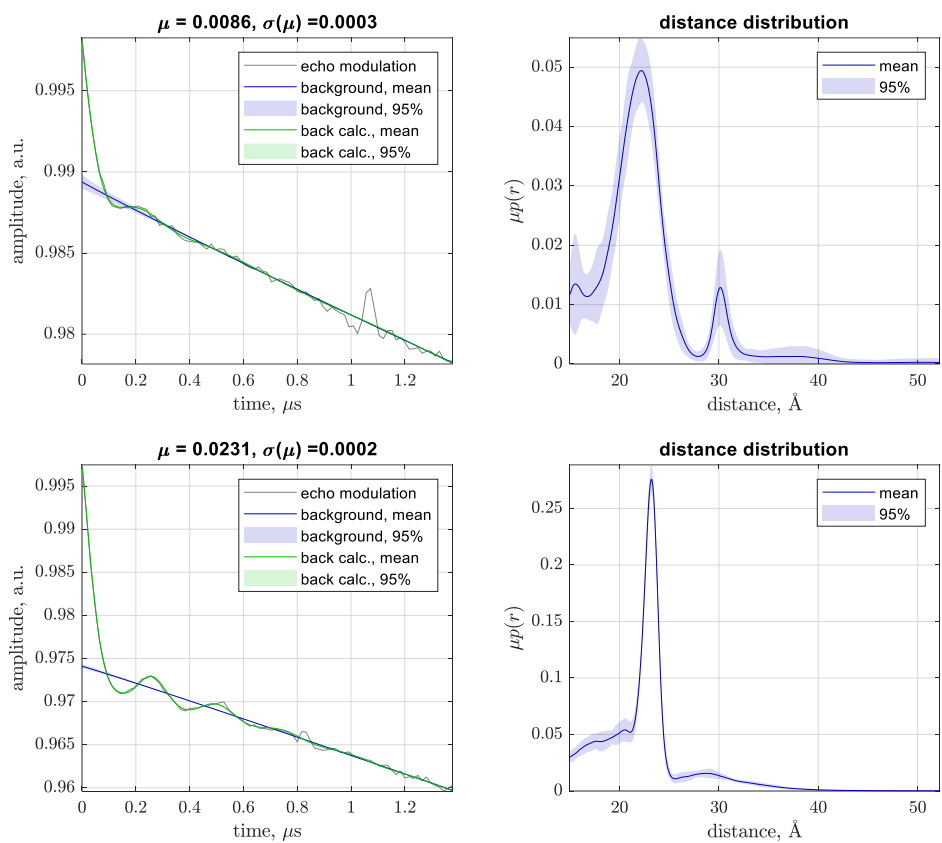

*Figure 10. An illustration of neural network resilience to running echo artefacts.* The time traces are DEER data for a bis-Gd(III)-DOTA model compound, measured with a large frequency separation (top row: 0.363 GHz, bottom row: 1.090 GHz) between the pump pulse and the observe pulse to reduce the effects of non-secular dipolar terms and zero-field splitting [44]. A running echo, ignored by the neural network, is visible at 1.1 μs (top left panel) and 0.8 μs (bottom left panel). The distance peaks at 30 Å and below 20 Å in the right panels are artefacts that depend on the frequency separation between pump and probe pulse; they arise because the kernel in Eq (13) had been derived for spin-1/2 particles, which Gd(III) is not.

Zero-field splitting makes quantum mechanics of high-spin pairs exceedingly difficult [42,43] – at the moment, there are no computationally affordable ways to process Gd(III) DEER data unless pulse frequencies are positioned so that the system stays in the weak coupling regime where Eq (13) is applicable. That approximation creates artefacts in the distance distribution. At the time of writing, they cannot be eliminated, but can be identified because their position and relative intensity fluctuate as a function of pump and probe pulse frequency separation (Figure 10, right panels).

### 8.4 Orientation selection

Microwave pulse widths in DEER and RIDME sequences are typically tens of nanoseconds long, corresponding to the excitation bandwidth in the tens of MHz. This is much less than the width of the *g*-tensor powder patterns of common spin labels (~ 500 MHz for nitroxides at W-band); the result is called orientation selection – some orientations of the spin system see the microwave pulse, others do not [78]. Because the kernel in Eq (13) is a full powder average, sending orientationally selected data into DEERNet is equivalent to sending corrupted data: dipolar modulation frequencies and depths strongly depend (Figure 11, left panel) on where the user decides to apply the pulse.

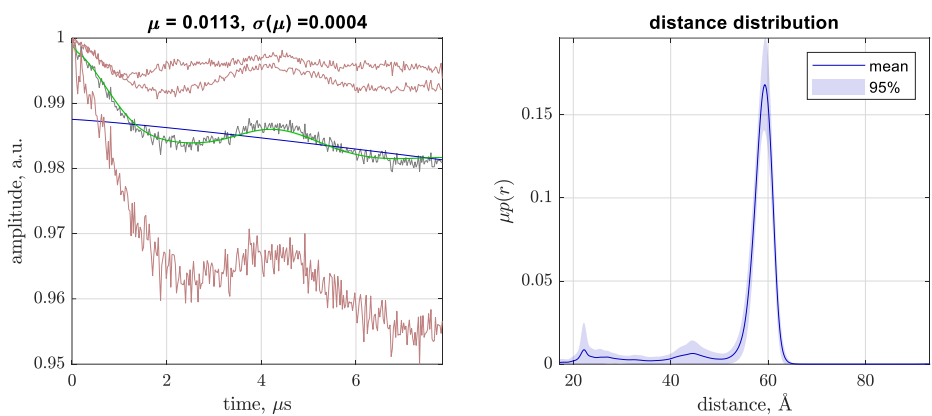

*Figure 11. An example of orientation selection in a 4-pulse DEER experiment. The sample is an ERp29 protein [79] dimer mutant (S114C) spin-labelled with Gd C1 and MTSL tags [80]. The orientation selection effect appears because the width of MTSL powder pattern at W-band is around 500 MHz, but the excitation bandwidth of the pump pulse is only ~10 MHz. The three red traces in the left panel have been recorded at three different pump frequencies inside the MTSL powder pattern; the grey trace with the background signal (blue line) is the average of the three red traces; the distance distribution on the right hand side was computed for that average.*

In situations where the powder pattern width is significantly larger than the pulse excitation width, DEER data must be acquired with pulse position averaging – several data sets with different pump and probe pulse frequencies must be acquired and averaged (Figure 11, grey trace on the left) to make sure that the isotropic powder average kernel in Eq (13) is applicable. It is not possible to train a neural network that would process arbitrary orientationally selected DEER / RIDME data because the one-to-one correspondence between the distance distribution and the modulation is broken.

## 9. Conclusions and outlook

In the practical testing (Sections S2-S4 in the Supplementary Information) performed by the groups involved in this study, deep neural network processing of DEER and RIDME data matches or exceeds the performance of the existing tools, insofar as "performance" may be defined for experimental datasets with uncertain ground truth; the same conclusion was reached for DEER in the recent community white paper [74]. The unique advantage of having no user-adjustable parameters, and thus no subjective bias, is pertinent because it removes the element of user discretion and error from the fiddly and subjective process of selecting regularisation parameters.

So far, common use scenarios for neural networks in DEER data processing have been: (a) in difficult cases where user bias in the choice of baseline and regularisation parameters must be eliminated; (b) as a second opinion alongside Tikhonov or model fitting tools; (c) as a source of initial guesses for Tikhonov or model fitting tools; (d) in situations when the users of DEER spectroscopy have no mathematical physics background and no interest in methodology, and only seek a reliable data processing tool. An appealing feature of *DEERNet* is that background signal retrofitting is optional: during distance distribution extraction, neural networks simply eliminate the background with a notch filter [30]; it may be retrofitted, if necessary, as described in Section 4.

The unexpectedly strong performance of neural network based Fredholm solvers merits a special mention – a recurring theme in the user feedback is the resilience of *DEERNet* to dismal experimental realities like low signal-to-noise ratios and incompletely sampled datasets. We would note here that many experimental data processing problems across natural (and unnatural) sciences reduce to solving a Fredholm equation; in this context, neural networks have a bright future.

### Acknowledgements

This work was funded by a grant from Leverhulme Trust (RPG-2019-048). The authors acknowledge the use of the IRIDIS High Performance Computing Facility, and associated support services at the University of Southampton. The deer drawing in the centre of Figure 1 was done by an anonymous artist with the pseudonym HellCorpCeo. Studentship funding and technical support from MathWorks are gratefully acknowledged. This research was supported by grants from NVIDIA and utilised NVIDIA Tesla A100 GPUs through the Academic Grants Programme. We also acknowledge funding from the Royal Society (University Research Fellowship for JEL) and EPSRC (EP/R513337/1 studentship for HR and EP/L015110/1 studentship for MJT). Author contributions were as follows: conceptualisation (IK, GJ), development of methodology (JK, TC, IK), programming and optimisation (JK, TC, IK), verification and validation (LG, EB, AF, DG, AE, LFI, GJ, HR, MJT, JEL, KK, MY), statistical analysis (JK, TC, IK), experimental data collection and annotation (LG, EB, AF, DG, AE, LFI, GJ, HR, MJT, JEL, KK, MY), code documentation and management (JK, TC, IK), paper writing and editing (all authors), oversight and coordination (IK), funding and computer equipment acquisition (IK).

### Statement of competing interests

All authors declare that they have no competing interests.

### Data and materials availability

All data and program code associated with this manuscript may be obtained by downloading *Spinach 2.7* or later versions from *https://spindynamics.org* web site.

# Neural networks in pulsed dipolar spectroscopy: a practical guide

Jake Keeley[1], Tajwar Choudhury[1], Laura Galazzo[2], Enrica Bordignon[2], Akiva Feintuch[3], Daniella Goldfarb[3], Hannah Russell[4], Michael J. Taylor[4], Janet E. Lovett[4], Andrea Eggeling[5], Luis Fabregas Ibanez[5], Katharina Keller[5], Maxim Yulikov[5], Gunnar Jeschke[5], Ilya Kuprov[1,*]

[1]School of Chemistry, University of Southampton, Southampton SO17 1BJ, United Kingdom.
[2]Faculty of Chemistry and Biochemistry, Ruhr University Bochum, Bochum, Germany.
[3]Department of Chemical Physics, Weizmann Institute of Science, Rehovot 7610001, Israel.
[4]SUPA School of Physics and Astronomy, and BSRC, University of St Andrews, North Haugh, St Andrews, KY16 9SS, United Kingdom.
[5]Department of Chemistry and Applied Biosciences, Swiss Federal Institute of Technology in Zurich, Vladimir Prelog Weg 2, CH-8093 Zürich, Switzerland.

*i.kuprov@soton.ac.uk

## S1. Sparsely sampled DEER experiments

In electron spin resonance, the four-pulse DEER pulse sequence consists of a refocused echo observer sequence [$\pi/2 - \tau_1 - \pi - \tau_1 + \tau_2 - \pi - \tau_2 - echo$] and a pump $\pi$ pulse that is applied during the inter-pulse delay of length $\tau_1 + \tau_2$. The first $\pi/2$ pulse generates transverse magnetization of the observer spin A that defocuses under the interactions of this spin with its environment during inter-pulse delay $\tau_1$. The following $\pi$ pulse inverts its phase, so that all interactions, including the dipole-dipole interaction are refocused after another delay $\tau_1$. This point corresponds to $t = 0$. In the absence of the pump pulse, the remaining subsequence $\tau_2$-$\pi$-$\tau_2$ causes another defocusing and refocusing of the same type. However, the pump $\pi$ pulse inverts spin B that is coupled to spin A. Thereby, the resonance frequency of spin A is changed by the dipole-dipole coupling $\omega_D \propto r^{-3}\left(3\cos^2\theta - 1\right)$. The second refocusing thus occurs at a phase difference $\pm \omega_D t$ with the sign depending on the initial state of spin B. Since the two states connected by the resonant B-spin transition are equally populated in the high-temperature approximation, the echo signal of a fully excited isolated spin pair varies as $\cos(\omega_D t)$. In typical biological applications, spins A and B are of the same type and only fractions of all spins are excited by the observer subsequence and pump pulse. Hence, only a fraction $\mu$ of the B spins is inverted. For A spins coupled to an unexcited B spin, the echo phase does not change. The signal of an isolated spin pair at a single orientation thus becomes $1 - \mu + \mu \cos(\omega_D t)$ with modulation depth $\mu$.

A sample of full length polypyrimidine tract binding protein 1 (PTBP1) in a complex with EMCV IRES RNA (domains D to F), prepared according to the protocol in [1], was kindly provided by Christoph Gmeiner. The protein was spin-labelled with MTSSL at residues S205C and Q352C. A solution of the protein-RNA-complex with 12.5 $\mu M$ protein concentration was prepared in a 1:1 (volume) mixture of a phosphate buffer (10 mM sodium phosphate, 20 mM NaCl, pH = 6.5 in $D_2O$) and $d_8$-glycerol. 30 µL of the solution were placed into a 3 mm outer diameter fused silica capillary and rapidly frozen by immersion into liquid nitrogen.

Uniformly and sparsely sampled DEER measurements were performed at 50 K using a commercial Q-band EPR spectrometer (Bruker ElexSys II E580 with a SpecJet II high-speed transient signal averaging module and a PatternJet II pulse programming module) equipped with a 200 W traveling wave tube amplifier and a home-built resonator for 3 mm outer diameter capillaries. Four-pulse DEER experiments $(\pi/2)_{obs} - \tau_1 - \pi_{obs} - t - \pi_{pump} - (\tau_1 + \tau_2 - t) - \pi_{obs} - \tau_2$ were used, with pulse durations of $t_{\pi/2,obs} = t_{\pi,obs} = 16$ ns and $t_{\pi,pump} = 12$ ns, and pulse frequencies of 34.390 GHz (pump, the maximum of the nitroxide spectrum) and 34.490 GHz (observe). 8-step nuclear modulation averaging was used in all cases. Inter-pulse delays $\tau_1$ and $\tau_2$ were set to 400 ns and 8 µs, respectively and the dead time delay was 280 ns. All measurements were performed using 5 shots per point, 6120 µs shot repetition time, and 50 scans.

Uniformly sampled measurements were performed using default PulseSPEL scripts (available in the libraries supplied with Bruker EPR spectrometers) for the 4-pulse DEER experiment. Pump pulse timings followed a 327-point uniform grid with $\Delta t = 24$ ns and $t_{max} = 8$ µs. Acquisition was completed in 2.3 hours. Sparsely sampled measurements were performed using custom PulseSPEL and ProDeL scripts enclosed with this Supplementary Information and based on the algorithm described in [2].

Pump pulse timings were distributed non-uniformly according to $\exp(-5t/t_{max})$ sample density function. Sparsely sampled measurements were performed at three sampling densities: 50% (9.25 hours measurement time), 25% (5.25 hours), and 10% (2.65 hours). The increase in measurement time, despite the reduced number of acquired points is a consequence of the overhead in the execution of acquisition scripts in ElexSys II E580, described in detail in [2]; the manufacturer has committed to resolving firmware problems causing those overheads. This problem does not arise with home-built instruments.

## S2. Validation run on published DEER datasets

This section contains DEERNet outputs for two diverse and *de facto* standard test data sets: one distributed with DEERAnalysis [3] and the other included with our previous paper dealing with non-methodological aspects of DEERNet [4]. The primary data may be downloaded as a part of DEERAnalysis (https://epr.ethz.ch/software.html) and Spinach (https://spindynamics.org) respectively. Biochemical, technical, and methodological information associated with these datasets is available in the references cited in figure captions.

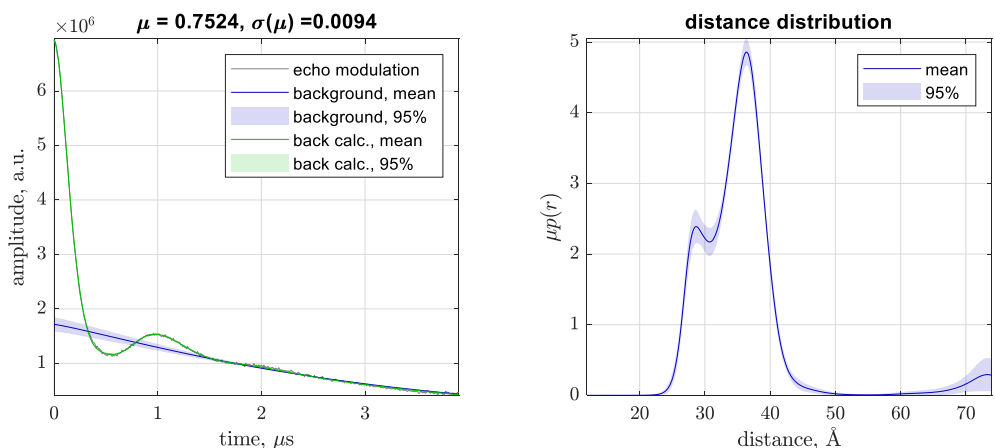

*Figure S1. Validation run using a data set (deer_tri_36_50K) supplied with DEERAnalysis [3]. The 13 datasets used here for validation are included with Spinach and DEERAnalysis example sets; they are designed to cover the reasonable use cases of organic and biomolecular DEER spectroscopy.*

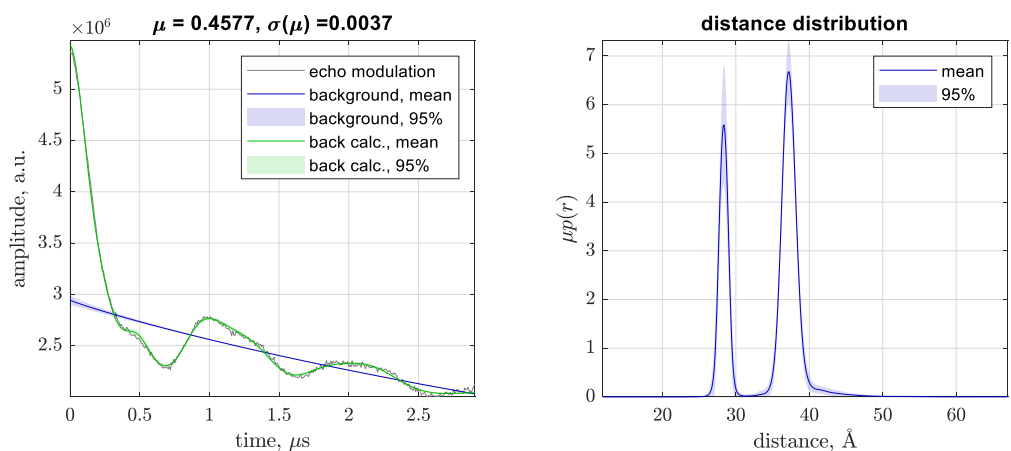

*Figure S2. Validation run using a data set (deer_mixture_80K_m_8_5scans) supplied with DEERAnalysis [3]. The 13 datasets used here for validation are included with Spinach and DEERAnalysis example sets; they are designed to cover the reasonable use cases of organic and biomolecular DEER spectroscopy.*

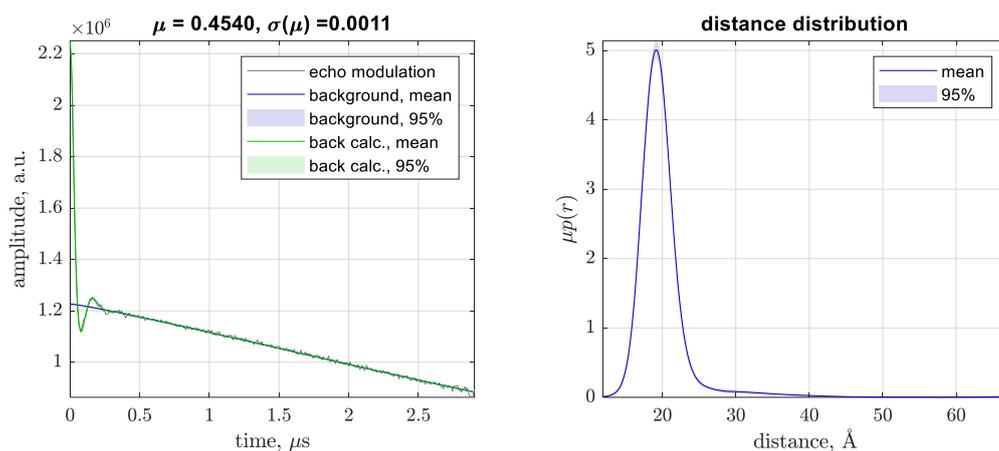

*Figure S3. Validation run using a data set (deer_bi_oligo_n8_50K) supplied with DEERAnalysis [3].* The 13 datasets used here for validation are included with Spinach and DEERAnalysis example sets; they are designed to cover the reasonable use cases of organic and biomolecular DEER spectroscopy.

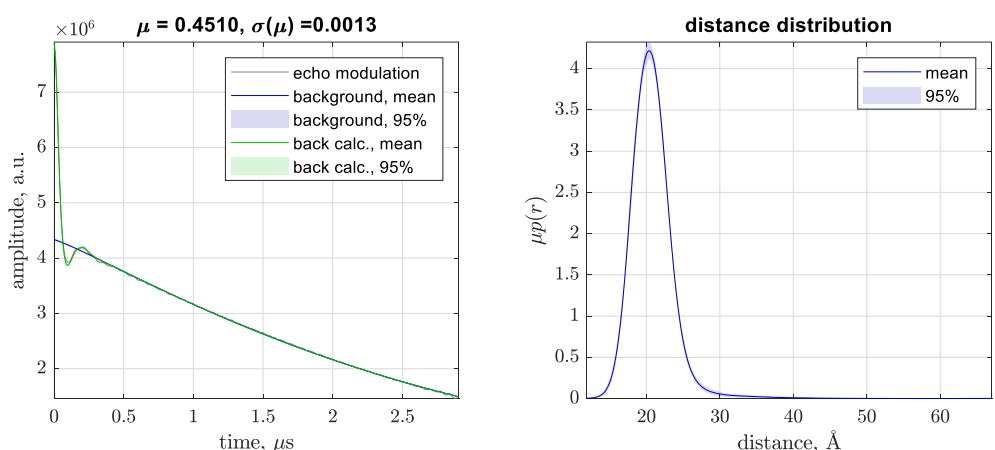

*Figure S4. Validation run using a data set (deer_bi_oligo_n10_50K) supplied with DEERAnalysis [3].* The 13 datasets used here for validation are included with Spinach and DEERAnalysis example sets; they are designed to cover the reasonable use cases of organic and biomolecular DEER spectroscopy.

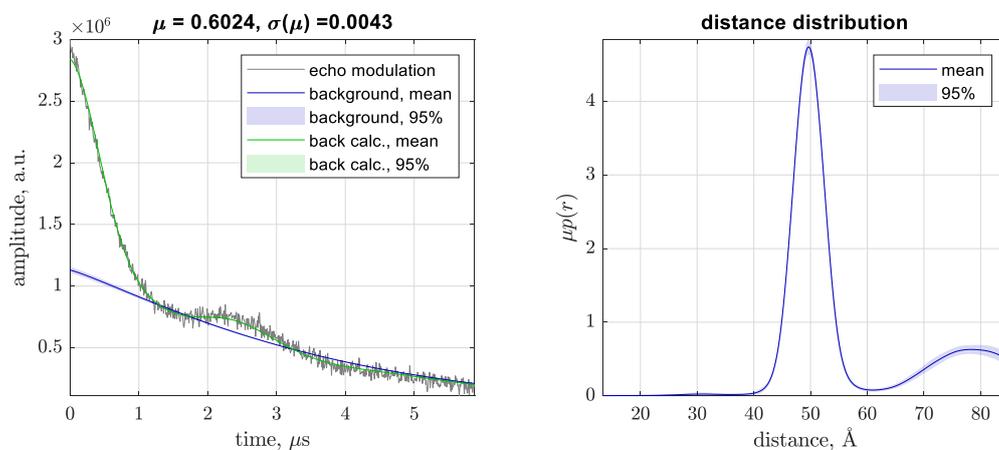

*Figure S5. Validation run using a data set (deer_bi_50_50K) supplied with DEERAnalysis [3].* The 13 datasets used here for validation are included with Spinach and DEERAnalysis example sets; they are designed to cover the reasonable use cases of organic and biomolecular DEER spectroscopy.

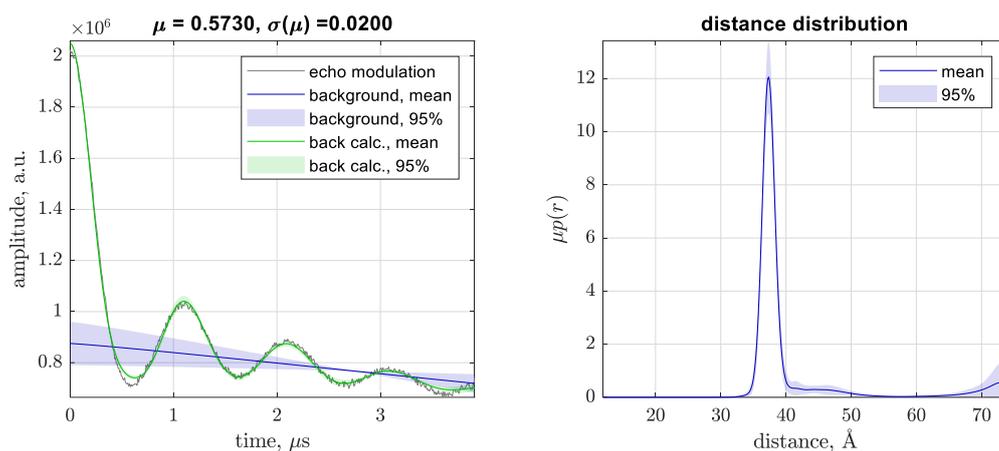

*Figure S6. Validation run using a data set (deer_bi_36_50K) supplied with DEERAnalysis [3].* *The 13 datasets used here for validation are included with Spinach and DEERAnalysis example sets; they are designed to cover the reasonable use cases of organic and biomolecular DEER spectroscopy.*

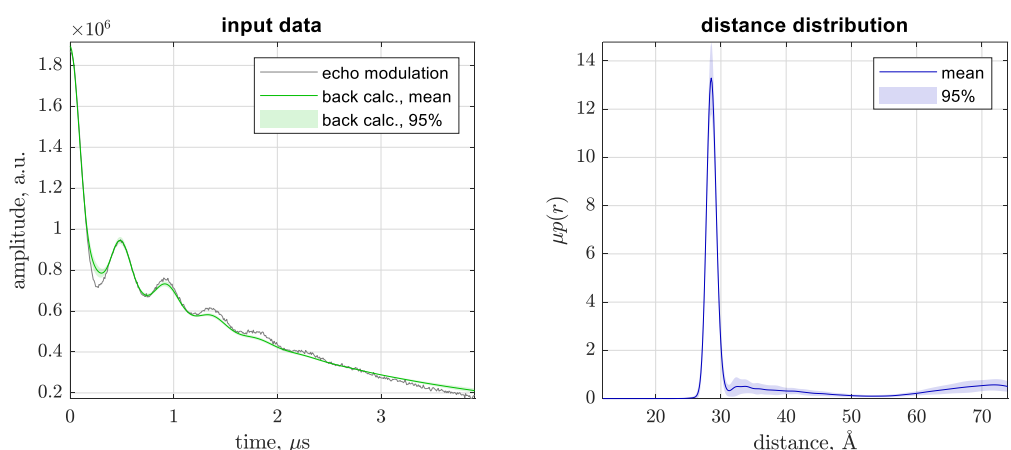

*Figure S7. Validation run using a data set (deer_bi_28_50K) supplied with DEERAnalysis [3].* *The 13 datasets used here for validation are included with Spinach and DEERAnalysis example sets; they are designed to cover the reasonable use cases of organic and biomolecular DEER spectroscopy.*

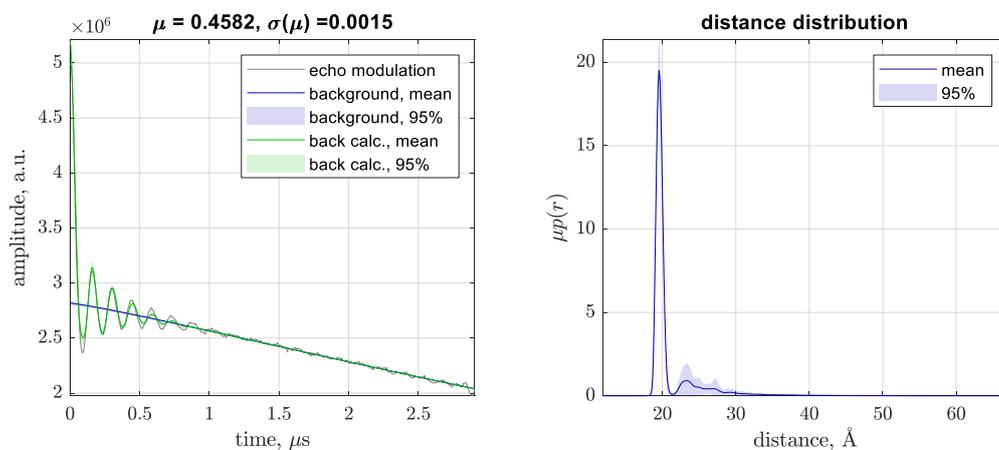

*Figure S8. Validation run using a data set (deer_bi_19_50K) supplied with DEERAnalysis [3].* *The 13 datasets used here for validation are included with Spinach and DEERAnalysis example sets; they are designed to cover the reasonable use cases of organic and biomolecular DEER spectroscopy.*

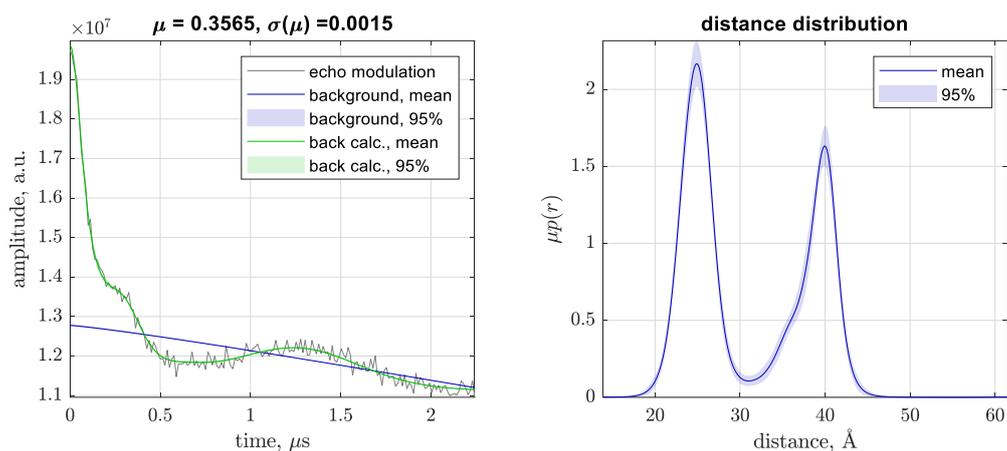

*Figure S9. Validation run using a data set (deer_252cl_113scans) supplied with DEERAnalysis [3].* The 13 datasets used here for validation are included with Spinach and DEERAnalysis example sets; they are designed to cover the reasonable use cases of organic and biomolecular DEER spectroscopy.

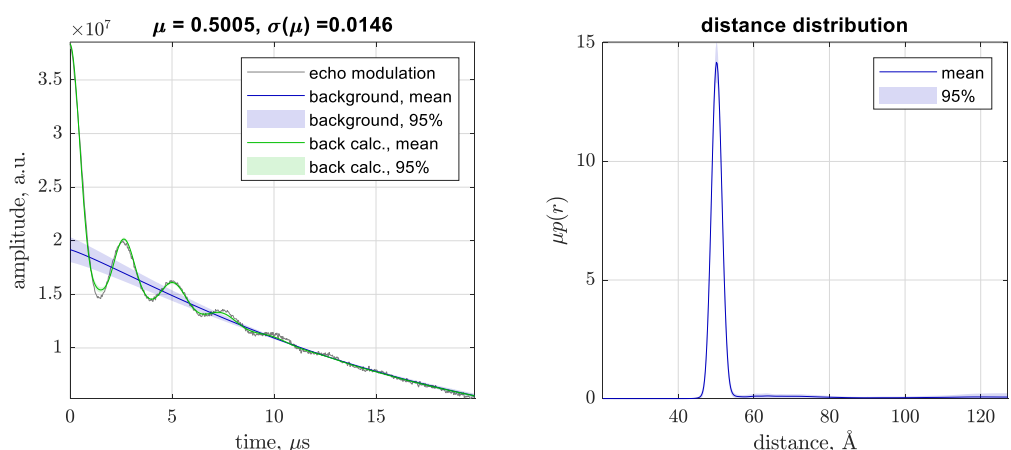

*Figure S10. Validation run using a data set (dOTP_5nm) supplied with DEERAnalysis [3].* The 13 datasets used here for validation are included with Spinach and DEERAnalysis example sets; they are designed to cover the reasonable use cases of organic and biomolecular DEER spectroscopy.

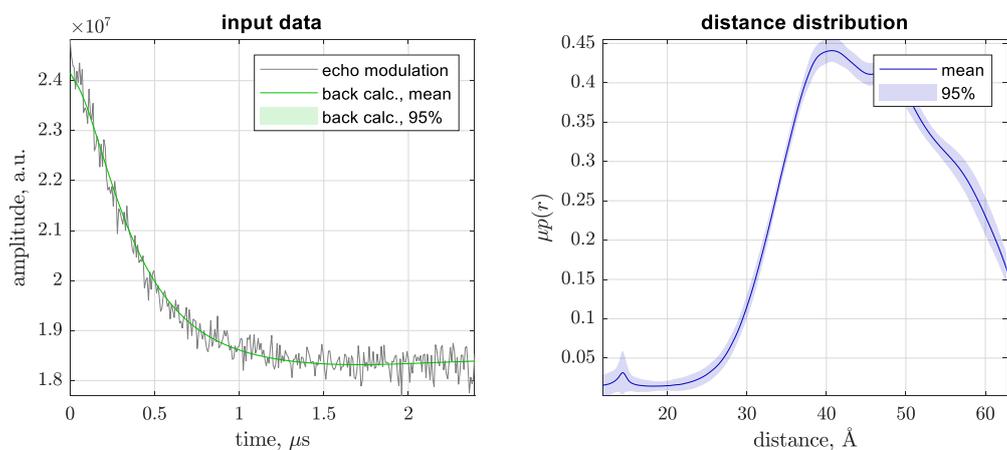

*Figure S11. Validation run using a data set (CT_deer_broad) supplied with DEERAnalysis [3].* The 13 datasets used here for validation are included with Spinach and DEERAnalysis example sets; they are designed to cover the reasonable use cases of organic and biomolecular DEER spectroscopy.

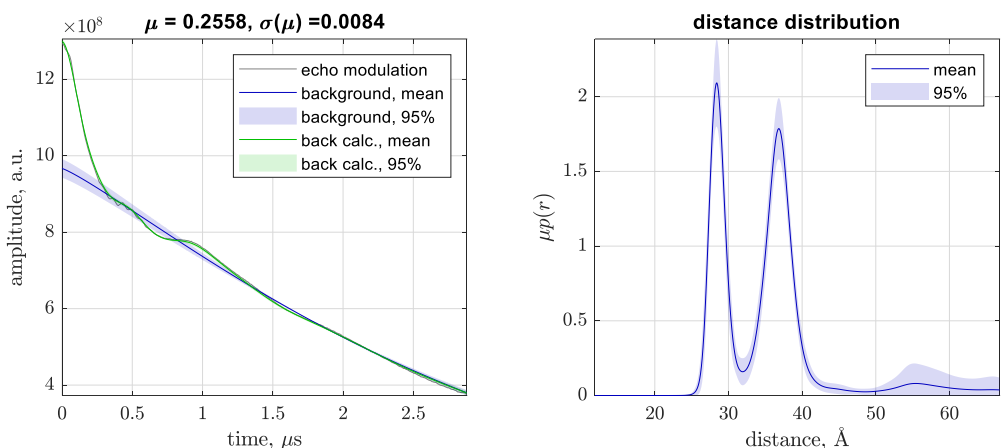

*Figure S12. Validation run using a data set (CT_DEER_mix_28_36) supplied with DEERAnalysis [3].* The 13 datasets used here for validation are included with Spinach and DEERAnalysis example sets; they are designed to cover the reasonable use cases of organic and biomolecular DEER spectroscopy.

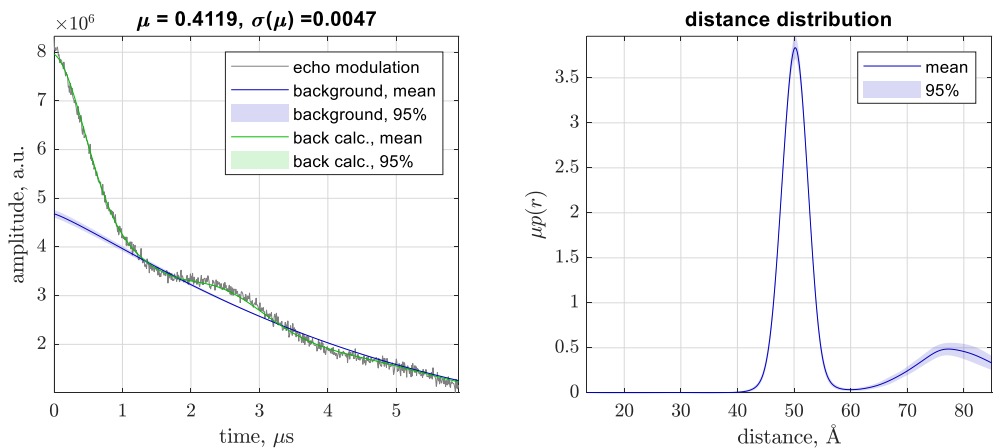

*Figure S13. Validation run using a data set (CT_DEER_5nm) supplied with DEERAnalysis [3].* The 13 datasets used here for validation are included with Spinach and DEERAnalysis example sets; they are designed to cover the reasonable use cases of organic and biomolecular DEER spectroscopy.

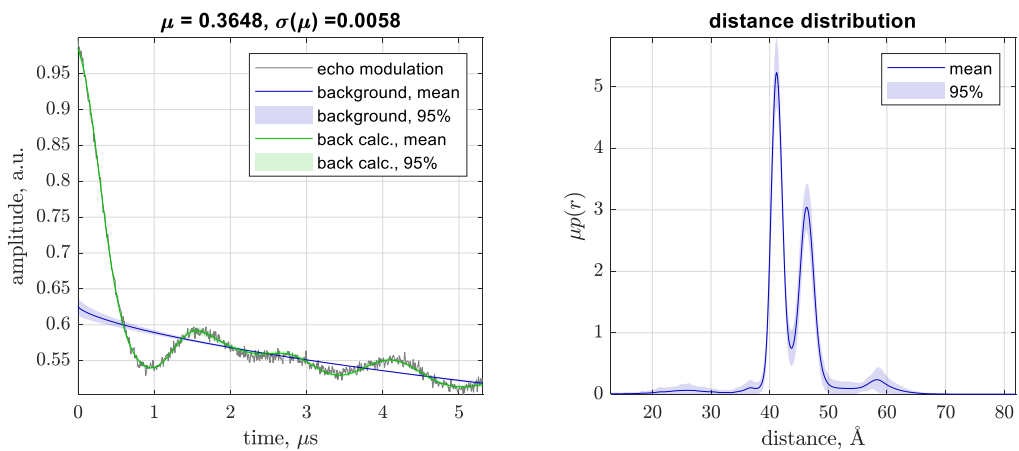

*Figure S14. Sample I from the DEERNet paper [4] by Worswick et al.* The sample is a site pair V96C/I143C in the lumenal loop of a double mutant of light harvesting complex II, with iodocateamido-PROXYL spin labels attached to the indicated cysteines [5].

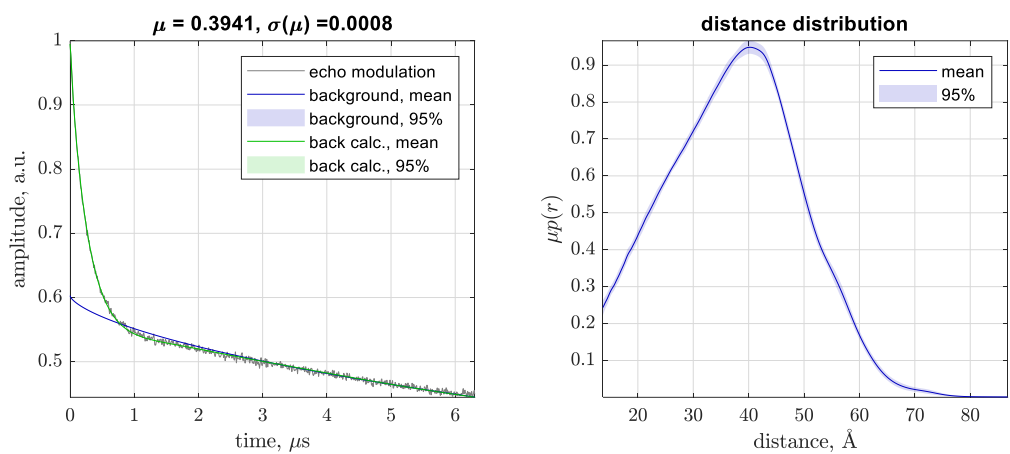

*Figure S15. Sample II from the DEERNet paper [4] by Worswick et al. The sample is a site pair S3C/S34C in the N-terminal domain of a double mutant of the light harvesting complex II, with iodoacetamido-PROXYL spin labels attached to the indicated cysteines [5].*

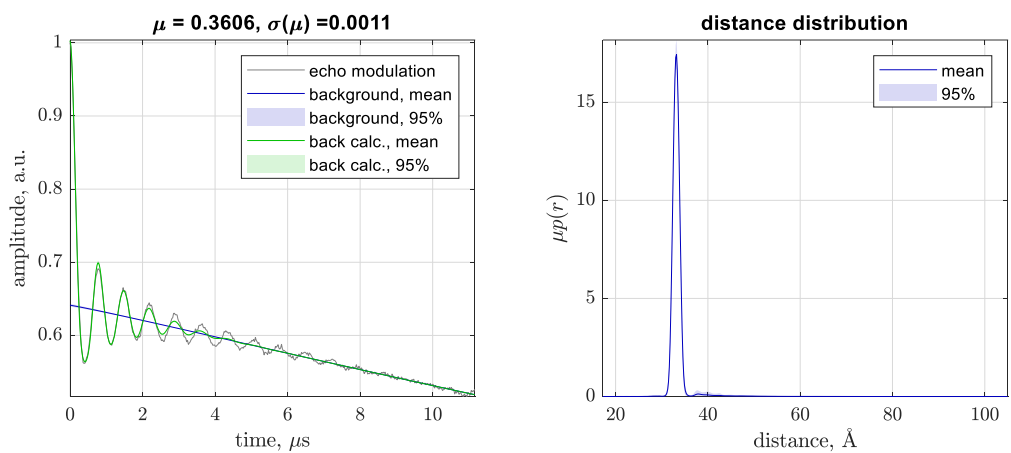

*Figure S16. Sample III from the DEERNet paper [4] by Worswick et al. The sample is end-labelled oligo-(para-phenyleneethynylene), a rigid linear molecule described as compound 3a in [6].*

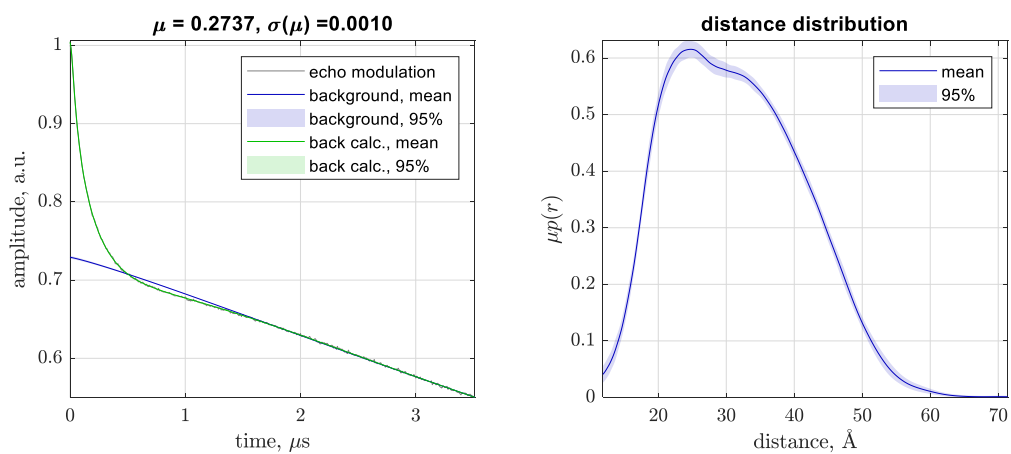

*Figure S17. Sample IV from the DEERNet paper [4] by Worswick et al. The sample is [2]catenane (a pair of large interlocked ring molecules) with a nitroxide spin label on each ring [7].*

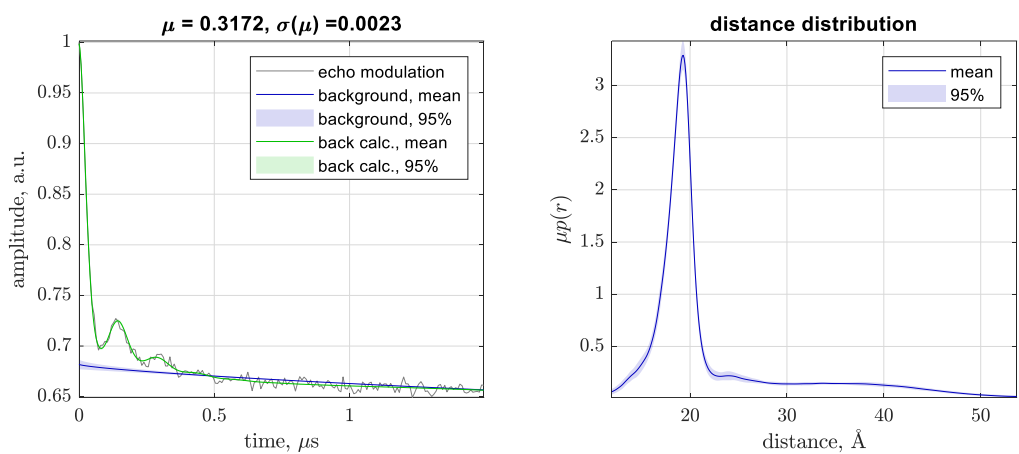

*Figure S18. Sample V from the DEERNet paper [4] by Worswick et al.* The signal is generated by pairs of nitroxide radicals tethered to the surface of gold nanoparticles [8].

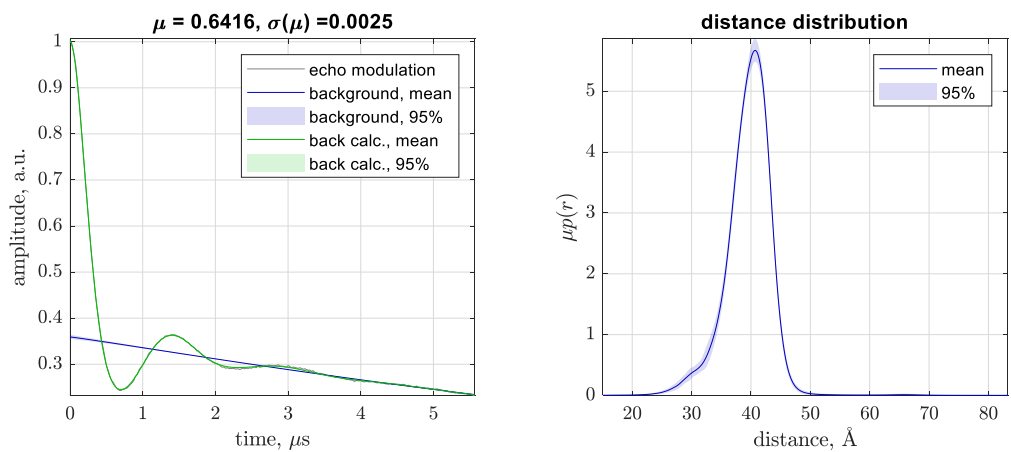

*Figure S19. Sample VI from the DEERNet paper [4] by Worswick et al.* The sample is a rigid molecular triangle labelled with nitroxide radicals on two corners out of three [9].

## S3. Validation run on published RIDME datasets

This section contains DEERNet outputs for RIDME datasets recorded for a protein homodimer of copper amine oxidase from *Arthrobacter globiformis* (AGAO) containing a $Cu^{2+}$ ion bound to a surface site on each monomer and one MTSL spin label per monomer [10] (Figures S31-S33), and for a $Cu^{2+}$-$Cu^{2+}$ pair with copper ions coordinated by PyMTA ligands (Figure S34) connected by a rigid oligo-phenyleneethinylene backbone [11]. The original data may be downloaded as a part of *Spinach* (https://spindynamics.org). Biochemical, technical, and methodological information associated with these datasets is available in the references cited in figure captions.

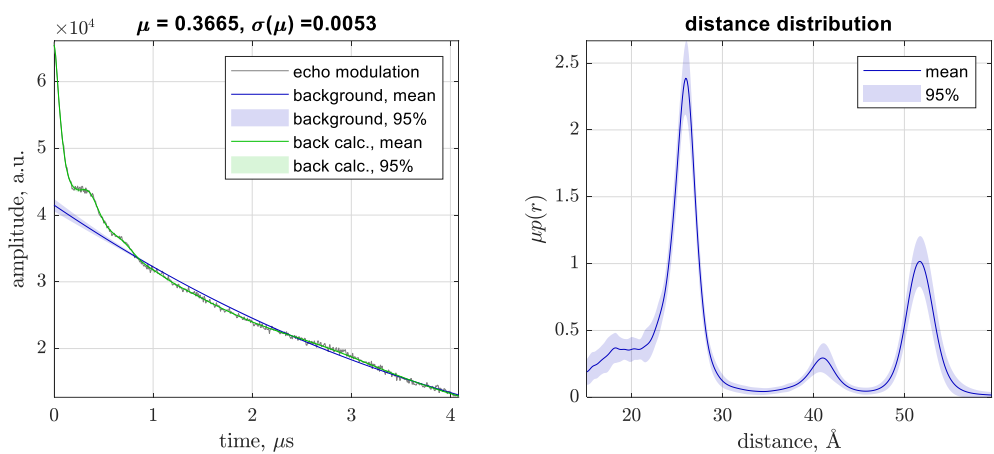

*Figure S20. RIDME experiment on the AGAO+Cu sample from [10].* The sample is a nitroxide-labelled, copper-bound AGAO homodimer. RIDME measurements were performed with a mixing time of 5 μs between a nitroxide spin-label and bound copper(II) on the same and opposite monomers. Further experimental details may be found in the reference cited.

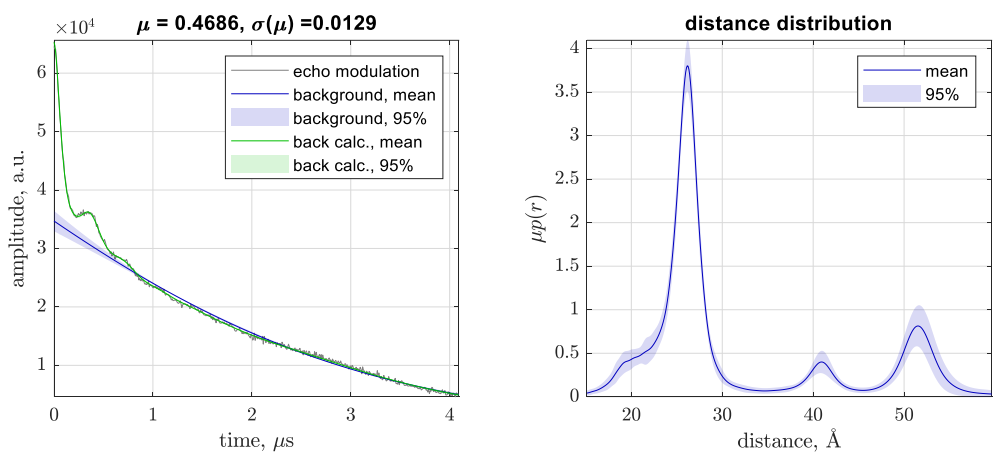

*Figure S21. RIDME experiment on the AGAO+Cu sample from [10].* The sample is a nitroxide-labelled, copper-bound AGAO homodimer. RIDME measurements were performed with a mixing time of 10 μs between a nitroxide spin-label and bound copper(II) on the same and opposite monomers. The longer mixing time than for the data in Figure S31 acts to increase the overall background decay, and therefore make the second distance at 52 Angstroms more difficult to detect. Further experimental details may be found in the reference cited.

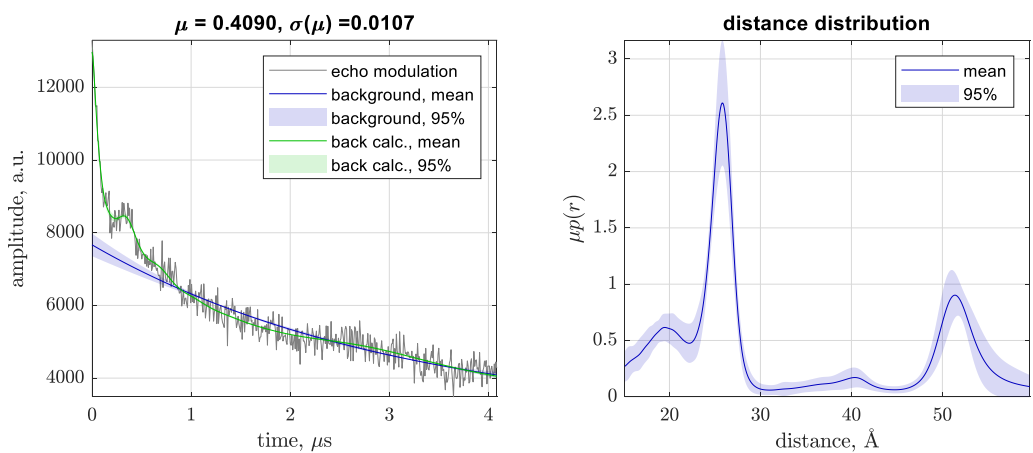

*Figure S22. RIDME experiment on the AGAO+Cu sample from [10].* The sample is a nitroxide-labelled, copper-bound AGAO homodimer wherein the $Cu^{2+}$ concentration was doubled relative to the sample in Figures S31 and S32, and the protein solution was diluted by a factor of 10. RIDME measurements were performed with a mixing time of 5 μs between a nitroxide spin-label and bound $Cu^{2+}$ on the same and opposite monomers. Further experimental details may be found in the reference cited.

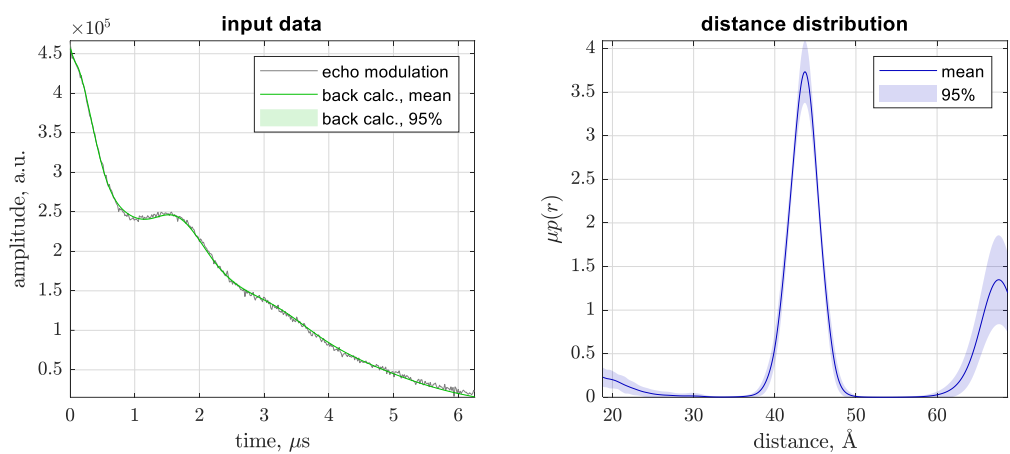

*Figure S34. Orientation averaged RIDME experiment on Cu(II)-PyMTA sample from [11].* The sample is a Cu(II)-Cu(II) spin pair with copper ions coordinated by PyMTA ligands connected by a rigid oligo-phenyleneethinylene backbone. This is an example of a significant distance peak being present at the long edge. In second generation DEERNet, this prevents background analysis from being run and triggers a console warning message to the user.

## S4. DEERNet vs. two-stage Tikhonov processing of RIDME data

This section contains a further illustration of the difference in the confidence level between DEERNet and two-stage Tikhonov methods for longer distance RIDME peaks. The sample is a protein homodimer of copper amine oxidase from *Arthrobacter globiformis* (AGAO) containing a $Cu^{2+}$ ion bound to a surface site on each monomer and one MTSL spin label per monomer [10]. The original data may be downloaded as a part of *Spinach* (https://spindynamics.org). Biochemical, technical, and methodological information associated with these datasets is available in the references cited in figure captions.

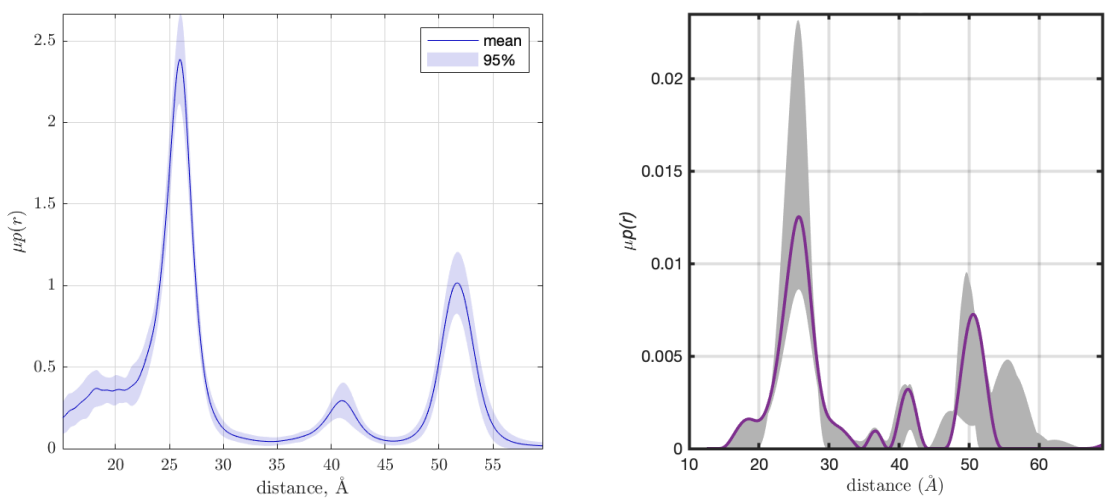

*Figure S35. Comparison of single-stage DEERNet processing (left) and two-stage Tikhonov regularised processing (right) of RIDME data from the AGAO+Cu sample from [10].* The sample is a nitroxide-labelled, copper-bound AGAO homodimer. RIDME measurements were performed with a mixing time of 5 µs and $\tau_2$ = 4400 ns between a nitroxide spin-label and bound copper(II) on the same and opposite monomers. Further experimental details may be found in the reference cited. The two-stage processing using DeerAnalysis involved background elimination with a stretched exponential model fit followed by regularised fitting of the form factor with respect to the distance distribution. The fitted background had a start value of 784 ns, a zero time of 137 ns, and a stretch parameter of 3.33; the background corrected trace was fitted using Tikhonov regularisation (parameter value of 316 from the L-curve criterion). Validation was done by incrementing the background start value between 5% and 50% of the length of the trace in 16 steps.

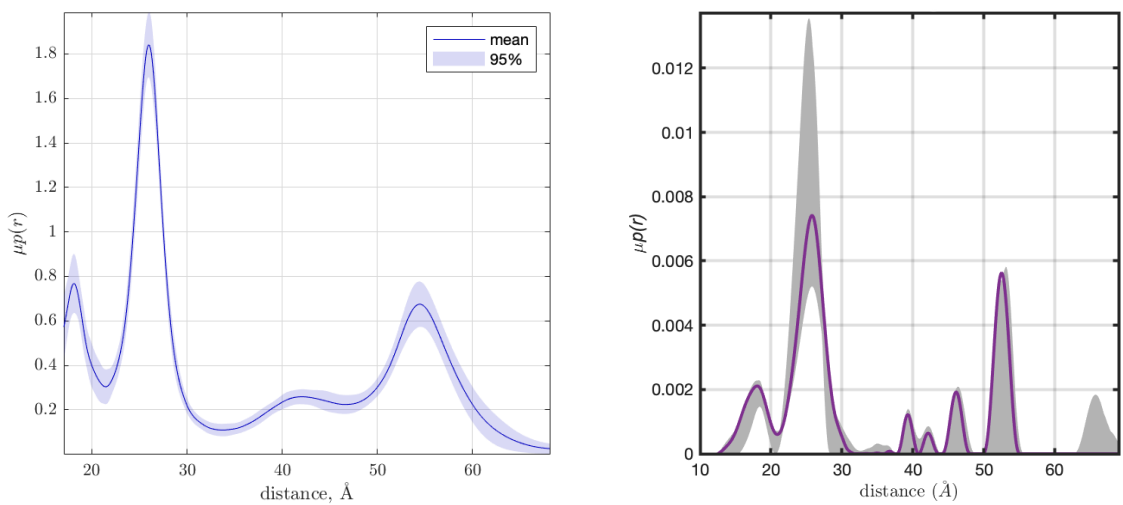

*Figure S36. Comparison of single-stage DEERNet processing (left) and two-stage Tikhonov regularised processing (right) of RIDME data from the AGAO+Cu sample from [10].* The sample is a nitroxide-labelled, copper-bound AGAO homodimer. RIDME measurements were performed with a mixing time of 5 μs and $τ_2$ = 6280 ns between a nitroxide spin-label and bound copper(II) on the same and opposite monomers. Further experimental details may be found in the reference cited. The two-stage processing using DeerAnalysis involved background elimination with a stretched exponential model fit followed by regularised fitting of the form factor with respect to the distance distribution. The fitted background had a start value of 2216 ns, a zero time of 135 ns, and a stretch parameter of 2.37; the background corrected trace was fitted using Tikhonov regularisation (parameter value of 316 from the L-curve criterion). Validation was done by incrementing the background start value between 5% and 40% of the length of the trace in 16 steps.